%% file: main.tex
\documentclass[journal=jctcce,manuscript=article,super=false]{achemso}
\usepackage[version=3]{mhchem} 
\usepackage{graphicx}
\usepackage{dcolumn}
\usepackage{color}
\usepackage{latexsym,bm}
\usepackage[normalem]{ulem}
\usepackage{multirow}
\usepackage{appendix}
\usepackage{amsmath}
\usepackage{ulem}
\usepackage{booktabs} 

\input{newcommands}

\title{Real-time time-dependent density functional theory simulations with range-separated hybrid functionals for periodic systems}
\author{Yuyang Ji}
\affiliation{Laboratory of Quantum Information, University of Science and
  Technology of China, Hefei 230026, Anhui,  China}

\author{Haotian Zhao}
\affiliation{Laboratory of Quantum Information, University of Science and
  Technology of China, Hefei 230026, Anhui, China}
  
\author{Peize Lin}
\affiliation{Institute of Artificial Intelligence, Hefei Comprehensive National Science Center, Hefei 230026, Anhui, China}

\author{Xinguo Ren}
\affiliation{Institute of Physics, Chinese Academy of Sciences, Beijing 100190, China}
\email{renxg@iphy.ac.cn}

\author{Lixin He}
\affiliation{Laboratory of Quantum Information, University of Science and Technology of
China, Hefei 230026, Anhui, China}
\alsoaffiliation{Institute of Artificial Intelligence, Hefei Comprehensive National Science Center, Hefei 230026, Anhui, China}
\email{helx@ustc.edu.cn}

\begin{tocentry}
 \centering
 \includegraphics[width=3.25in,height=1.75in]{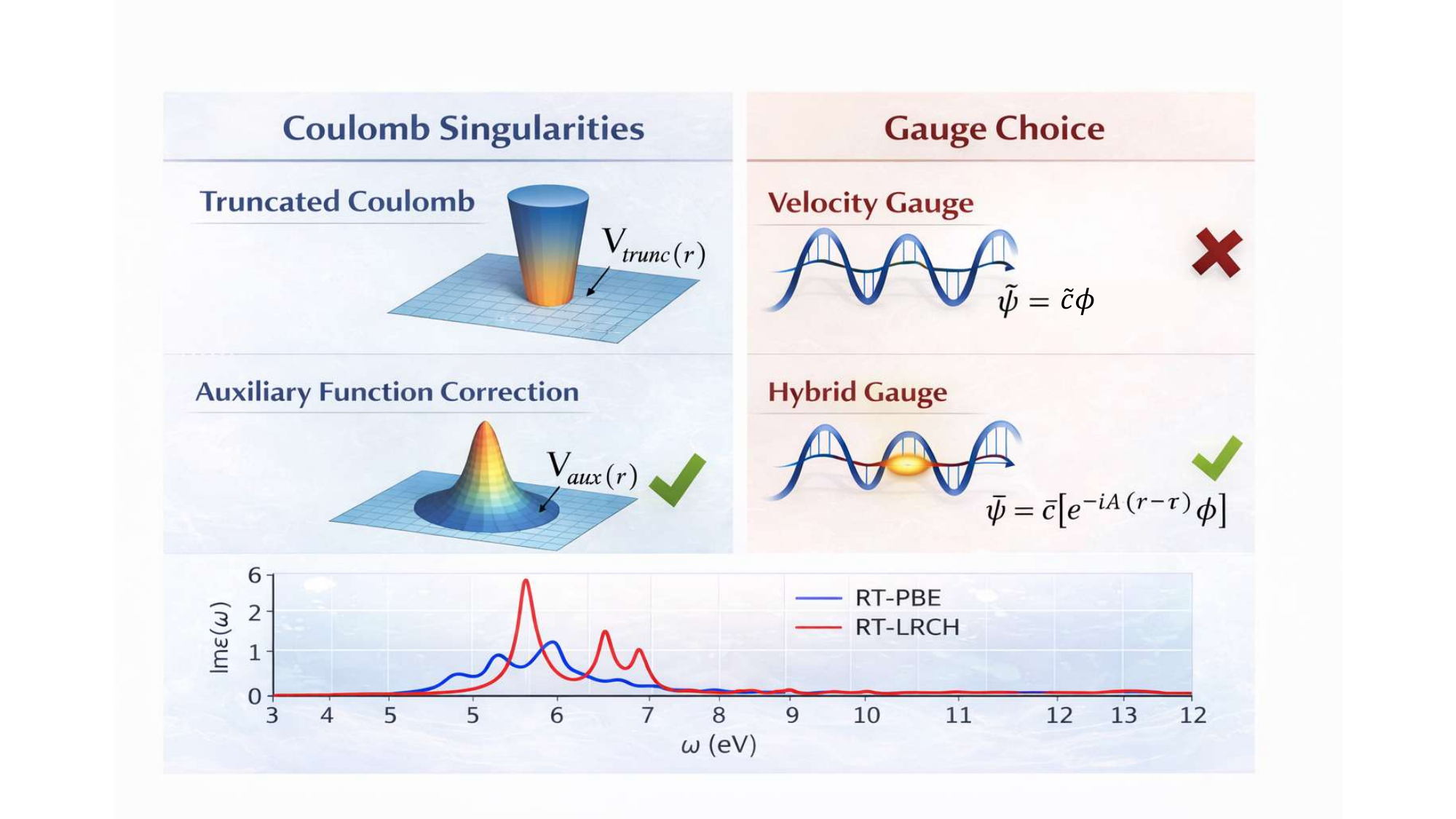}
\end{tocentry}

\begin{document}

\begin{abstract}
    Real-time time-dependent density functional theory (RT-TDDFT) is a powerful approach for investigating various ultrafast phenomena in materials. However, most existing RT-TDDFT studies rely on adiabatic local or semi-local approximations, which suffer from several shortcomings, including the inability to accurately capture excitonic effects in periodic systems. Combining RT-TDDFT with range-separated hybrid (RSH) functionals has emerged as an effective strategy to overcome these limitations. The RT-TDDFT-RSH implementation for periodic systems requires careful treatment of the Coulomb singularity and choosing proper gauges for the incorporation of external fields. We benchmark two schemes for treating the Coulomb singularity — the truncated Coulomb potential and the auxiliary-function correction — and find that the latter shows better convergence behavior and numerical stability for long-range corrected hybrid functions. Additionally, we assess the impact of gauge choice in simulations using numerical atomic orbitals and show that the recently proposed hybrid gauge incorporating position-dependent phases provides a more accurate description of excitonic absorption than the conventional velocity gauge. Our implementation significantly improves the accuracy of RT-TDDFT-RSH for modeling ultrafast excitonic dynamics in periodic systems.


\end{abstract}
\maketitle

\section{Introduction}
Ultrafast dynamics in condensed matter systems involve nonequilibrium phenomena triggered by femtosecond-scale electromagnetic excitation. These processes, which occur on femtosecond to picosecond timescales, are governed by the complex interplay of electronic, lattice, and spin degrees of freedom \cite{Exp/TDDFT/Gort/2018,Exp/TDDFT/Zahn/2021,Exp/TDDFT/Filippetto/2022,Exp/TDDFT/Zhang/2022,Exp/TDDFT/Koya/2023,Exp/TDDFT/Zhang/2023,Exp/TDDFT/Zou/2023,Exp/TDDFT/Vinci/2025}, leading to rapid changes and significant deviations from equilibrium. Such deviations can manifest as transient metallic states \cite{Exp/TDDFT/Medvedev/2023,Exp/TDDFT/Duan/2023}, symmetry-breaking transitions \cite{Exp/TDDFT/Li/2021,Exp/TDDFT/Zhou/2021,Exp/TDDFT/Sirica/2022}, demagnetization \cite{Exp/TDDFT/Zhou/2024,Exp/TDDFT/Mishra/2022,Exp/TDDFT/Chekhov/2021}, and energy transfer \cite{Exp/TDDFT/Yang/2021,Exp/TDDFT/Chen/2022}. Developing a quantitatively accurate theoretical framework to describe these dynamics is crucial for uncovering the underlying microscopic mechanisms and for guiding future experimental and technological advancements.

Real-time time-dependent density functional theory (RT-TDDFT) has emerged as a powerful first-principles approach for investigating ultrafast nonequilibrium dynamics \cite{TDDFT/Runge/1984,TDDFT/Gross/1990,TDDFT/Yabana/1999,TDDFT/Castro/2004,TDDFT/Otobe/2008,TDDFT/Gross/2012}. In typical RT-TDDFT simulations, time propagation is performed using wavefunctions obtained within the local-density approximation (LDA) or generalized gradient approximations (GGAs) \cite{TDDFT/APP/Guo/2025,TDDFT/APP/Kim/2022,TDDFT/APP/Li/2024,TDDFT/APP/Nie/2023,TDDFT/APP/Sone/2023,TDDFT/APP/Wu/2024}. However, the inherent delocalization errors of LDA and GGAs — notorious for limiting the accuracy of ground-state properties such as band gaps and potential energy surfaces \cite{Delocalization/Perdew/1982,Delocalization/Perdew/1983,Delocalization/Aron/2008,Delocalization/Bryenton/2022,Delocalization/Chen/2017,Delocalization/Dieter/2004,Delocalization/Hait/2018}—also compromise their performance in excited-state dynamics. These shortcomings are particularly evident in their inaccurate description of Rydberg and charge-transfer excitations in finite systems, as well as exciton dynamics in solids \cite{TDDFT/Appel/2003, TDDFT/Casida/1998, LCPBE/Tawada/2004, TDDFT/Sun/2021, TDDFT/Dreuw/2004, TDDFT/Kronik/2012}.

Hybrid density functionals (HDFs), which include a fraction of nonlocal Hartree-Fock exchange (HFX), can effectively reduce the delocalization errors. Within the generalized Kohn-Sham (GKS) framework \cite{GKS/Gorling/1997,GKS/Seidl/1996}, these functionals yield a non-local effective exchange-correlation (XC) potential, leading to significantly improved band gaps compared to the strict Kohn-Sham (KS) approach \cite{Basic/hohenberg/1964,Basic/kohn/1965}. Range-separated hybrid (RSH) functionals, in particular, can provide a more accurate description of the long-range asymptotic behavior of the XC potential, thereby considerably improving the prediction of excited-state properties when employed within the linear-response TDDFT (LR-TDDFT) framework \cite{LCPBE/Tawada/2004, TDDFT/Rohrdanz/2009, TDDFT/Kronik/2012, TDDFT/Refaely/2015, TDDFT/Ohad/2023, TDDFT/Liu/2020, TDDFT/Sun/2020}. However, the application of RSH functionals within the RT-TDDFT framework (denoted as RT-TDDFT-RSH hereafter) remains rather limited \cite{TDDFT/Pemmaraju/2019, TDDFT/Sun/2021, TDDFT/Shepard/2024}, mainly due to the high computational cost associated with the evaluation of the HFX, which formally scales as $O(N^4)$. In recent years, significant progress has been made in speeding up the HFX calculations. Most notably, the employment of the resolution-of-identity (RI) technique \cite{RI/Whitten/1973, RI/Dunlap/1979, RI/Florian/1998, RI/Vahtras/1993, RI/Dunlap/2010, RI/Ren/2012}, especially its localized variant
\cite{RI/Ihrig/2015, RI/Levchenko/2015, ABACUS/Lin/2020} within the numerical atomic orbital (NAO) basis set framework, drastically accelerates the evaluation of HFX, achieved by exploiting the sparsity of RI expansion coefficients and short-range nature of density matrix in real space. Such technological innovation enabled efficient calculations of hybrid functional energies, forces, and stresses in periodic systems containing hundreds to tens of thousands of atoms \cite{RI/Levchenko/2015, ABACUS/Lin/2021, RI/Lee/2022, RI/Bussy/2024, RI/Kokott/2024, ABACUS/Lin/2025}.  Given these advancements, it is now a timely endeavor to extend such efficient algorithms to the implementation of RT-TDDFT-RSH.

Nonetheless, implementing RT-TDDFT-RSH with NAOs is confronted with additional challenges. One key issue is the integrable divergence of the long-range Coulomb potential as $q \to 0$, which requires careful numerical treatment \cite{Singularity/Gygi/1986,Singularity/Massidda/1993,Singularity/Carrier/2007,Singularity/Spencer/2008,Singularity/Broqvist/2009,Singularity/Sundararaman/2013}. Two commonly used schemes are the truncated Coulomb potential \cite{Singularity/Spencer/2008,Singularity/Sundararaman/2013} and the auxiliary function correction method\cite{Singularity/Gygi/1986,Singularity/Massidda/1993,Singularity/Carrier/2007,Singularity/Broqvist/2009}. Both approaches work well for full-range Coulomb matrix, but the truncation scheme becomes problematic for long-range Coulomb matrix, within
the RI formulation for NAO basis sets, as will be shown later in this work.
Another crucial concern is the choice of gauge for describing external fields. As shown by \citet{TDDFT/Zhao/2025}, when expanding the Bloch wavefunctions in terms of atomic orbital (AO) basis sets, the commonly used velocity gauge lacks the necessary position-dependent phases in the expansion coefficients, leading to significant errors in RT-TDDFT simulations using the LDA/GGA approximations. This particularly affects current-related properties. The impact of this deficiency on RT-TDDFT-RSH simulations remains unexplored.

In this work, we first compare the performance of two distinct schemes for treating the Coulomb singularity in RSH functional calculations.
Benchmark results show that the auxiliary function correction method offers significant advantages for long-range corrected hybrid (LRCH) functionals.
Building upon these findings, we integrate RSH functionals into the RT-TDDFT framework and investigate the absorption spectra of bulk Si and monolayer hexagonal-BN (h-BN), under various gauge choices. We found that employing the hybrid gauge \cite{TDDFT/Zhao/2025}, which incorporates the important position-dependent phases, substantially improves the accuracy of exciton dynamics simulations. 
Finally, we demonstrate that our RT-TDDFT-RSH framework enables accurate and efficient exploration of excited-state properties in extended systems, as exemplified by its application to double perovskite material Cs$_2$NaInCl$_6$.

\section{Methodology}

In this section, we begin with a brief introduction to the RT-TDDFT equations and the gauge problem that arises when a time-dependent external potential is introduced. For periodic systems, integrating RSH functionals into this framework necessitates addressing the Coulomb singularity. We then discuss the explicit formulations for two regularization methods: the truncated Coulomb potential and the auxiliary function correction. This is followed by the concrete equations for RT-TDDFT-RSH with the appropriate hybrid gauge, as implemented within the NAO basis set framework; in particular, the HFX matrix within this specific gauge is presented. Finally, we develop a scheme for evaluating the current in the presence of nonlocal HFX by inserting a complete set of basis functions, enabling a consistent treatment within our formalism.


\subsection{RT-TDDFT and gauge transformation}
In RT-TDDFT, the central task is to solve the time-dependent KS (TDKS) equation:
\begin{equation}
i\frac{\partial}{\partial t}\psi_{n\bfk}(\bfr,t) = \hat{H}\,\psi_{n\bfk}(\bfr,t),
\end{equation}
where $\hat{H}$ and $\psi_{n\bfk}$ denote the Hamiltonian and the KS wave function, respectively \cite{TDDFT/Meng/2008}.  

Within a numerical atomic orbital (NAO) basis set $\{\phi_{i}(\bfr)\}$, the KS wave function can be expanded as \cite{ABACUS/Lin/2023/review}
\begin{equation}
    \psi_{n\bfk}(\bfr,t) = \frac{1}{\sqrt{N_\bfk}}\sum_{i}\sum_{\bfR} 
    c_{n\bfk}^i(t)\,e^{i\bfk\cdot\bfR}\,
    \phi_{i}(\bfr-\boldsymbol{\tau}_{i}-\bfR)\, 
\end{equation}
where $ c_{n\bfk}^i(t)$ are the time-dependent KS eigenvectors, $\boldsymbol{\tau}_{i}$ is the atomic position of the basis $i$ in the unit cell, $\bfR$ is a lattice vector, 
and $N_\bfk$ is the number of $\bfk$ points in the first Brillouin zone (BZ), equivalent to the number of lattice vectors ($N_\bfR$) in the Born-von K\'{a}rm\'{a}n supercell.
Substituting this expansion into the TDKS equation yields the matrix form of the equation in the NAO basis:
\begin{equation}
    i\sum_{j} S_{ij}(\bfk)\,
    \frac{\partial c_{n \bfk}^j(t)}{\partial t} =
    \sum_{j} H_{ij}(\bfk)\,c_{n \bfk}^j(t),
\end{equation}
where
\begin{equation}
H_{ij}(\bfk) = \sum_{\bfR} 
e^{i\bfk\cdot\bfR}\,
\bra{\phi_{i{\bf 0}}}\hat{H}\ket{\phi_{j \bfR}}, 
\quad
S_{ij}(\bfk) = \sum_{\bfR} 
e^{i\bfk\cdot\bfR}\,
\bracketw{\phi_{i {\bf 0}}}{\phi_{j \bfR}}
\end{equation}
are the Hamiltonian and overlap matrices, respectively, in the NAO basis. 
To introduce an external electric field, the simplest approach is to use the length gauge, in which the Hamiltonian is written as
\begin{equation}
\hat{H}_{L}(t) = \hat{H}_0 + \boldsymbol{E}(t) \cdot \bfr,
\end{equation}
where $\boldsymbol{E}(t)$ is a time-dependent electric field, and
\begin{equation}
    \hat{H}_{0} = -\frac{1}{2}\nabla^2 + V_{\rm H}(\boldsymbol{r}) +  V_{\rm xc} + V^{\rm ps}_{\rm L}(\boldsymbol{r}) + V^{\rm ps}_{\rm NL}
\end{equation}
is the Hamiltonian of the system in the absence of an external field.
Specifically, $V_{\rm H}$ and $V_{\rm xc}$ are the Hartree and exchange–correlation (XC) potentials, while $V^{\rm ps}_{\rm L}$ and $V^{\rm ps}_{\rm NL}$ denote the local and nonlocal components of the pseudopotential, respectively. We note that $V_{\rm xc}$ is local for LDA and GGA functionals, but non-local for HDFs.

Despite its simple form and straightforward implementation, the length gauge cannot be applied to periodic systems because the operator $\boldsymbol{E}(t)\cdot\bfr$ breaks the translational symmetry. 
To overcome this difficulty, the standard approach is to use the velocity gauge. 
In the velocity gauge, the generalized KS Hamiltonian is expressed as\cite{TDDFT/Pemmaraju/2018}
\begin{equation}
    \hat{H}_{V}(t) = -\frac{1}{2}\left(\nabla + \boldsymbol{A}(t)\right)^2 
    + V_{\rm H}(\boldsymbol{r})+ \tilde{V}_{\rm xc}
    + V^{\rm ps}_{\rm L}(\boldsymbol{r}) 
    + \tilde{V}^{\rm ps}_{\rm NL},
\end{equation}
where the vector potential is defined as 
$\boldsymbol{A}(t) = -\int_{0}^{t}\boldsymbol{E}(t')\,dt'$, 
and $\tilde{V}^{\rm ps}_{\rm NL} = e^{-i\boldsymbol{A}(t)\cdot\bfr}\,V^{\rm ps}_{\rm NL}\,e^{i\boldsymbol{A}(t)\cdot\bfrp}$ 
is the nonlocal pseudopotential in the velocity gauge. Special attention should be paid to the XC potential in
the velocity gauge: While 
$\tilde{V}_{\rm xc}=V_{\rm xc}$ for local and semilocal functionals, for HDFs with a portion of non-local Hartree-Fock exchange (HFX),  
\begin{equation}
\tilde{V}_{\rm xc}(\bfr,\bfrp, t)= \alpha
e^{-i\boldsymbol{A}(t)\cdot\bfr}\,V^{\rm HF}_{\rm x}(\bfr,\bfrp)\,e^{i\boldsymbol{A}(t)\cdot\bfrp} +
(1-\alpha)V_{\rm x}(\bfr)\delta(\bfr-\bfrp) + V_{\rm c}(\bfr)\delta(\bfr-\bfrp)\, ,
\end{equation}
where $V^{\rm HF}_{\rm x}$ and $V_{\rm x}$ denote the non-local and local/semilocal exchange potentials, respectively.
In essence, the nonlocal part of the exchange potential in HDFs follows a similar phase transformation as the nonlocal pseudopotentials.
However, while the velocity gauge is applicable to periodic systems, it suffers from accuracy issues arising from basis-set incompleteness in the NAO basis, as it ignores the phase variations in the NAOs, leading to significant errors in physical quantities such as current or energy \cite{TDDFT/Zhao/2025}. 

Recently, Zhao and He proposed a hybrid gauge approach~\cite{TDDFT/Zhao/2025}, 
which retains the efficiency and accuracy of the length gauge while remaining applicable to periodic systems, thereby overcoming the drawbacks of both the length and velocity gauges. 
In the hybrid gauge, the basis functions are modified by incorporating a time-dependent phase factor:
\begin{equation}
    \bar{\phi}_{i}(\bfr-\boldsymbol{\tau}_{i}-\bfR,t) 
    = e^{-i\boldsymbol{A}(t)\cdot(\bfr-\boldsymbol{\tau}_{i}-\bfR)} 
    \phi_{i}(\bfr-\boldsymbol{\tau}_{i}-\bfR).
\end{equation}
To distinguish it from the original NAOs, we add a bar symbol. 
Other quantities in the hybrid gauge are likewise denoted with a bar to distinguish them from those in other gauges.

Within this basis set, the Hamiltonian matrix is rewritten as
\begin{equation}
    \bar{H}_{i j}(\boldsymbol{k}, t) = \sum_{\boldsymbol{R}}
    e^{-i \boldsymbol{A}(t) \cdot \boldsymbol{\tau}_{i {\bf 0}, j \bfR}} \,
    e^{i \boldsymbol{k} \cdot \boldsymbol{R}}
    \langle \phi_{i {\bf 0}} \,|\, \hat{H}_0 + \boldsymbol{E}(t) \cdot 
    (\bfr-\boldsymbol{\tau}_{j}-\boldsymbol{R}) \,|\, \phi_{j \bfR} \rangle,
    \label{eq:Ham_hybrid_gauge}
\end{equation}
where $\boldsymbol{\tau}_{i {\bf 0}, j \bfR} = \boldsymbol{\tau}_{i} - \boldsymbol{\tau}_{j} - \boldsymbol{R}$ denotes the relative distance between the two atoms associated with the bra and ket states.  
Apart from the additional phase factor, the Hamiltonian remains essentially the same as in the length gauge, allowing it to retain the computational efficiency of the length gauge.  
Although a scalar potential is introduced, unlike in the length gauge, this potential is periodic in form, enabling the Hamiltonian to be applied to periodic systems.  
Benchmark results demonstrate that this approach effectively resolves the accuracy issues inherent to the velocity gauge~\cite{TDDFT/Zhao/2025}.

\subsection{The HFX matrix in the hybrid gauge}\label{sec:HFX_matrix}
Previously, the hybrid gauge scheme has been successfully applied to RT-TDDFT with GGA functionals \cite{TDDFT/Zhao/2025}. The major purpose of the present work is to extend this scheme to RT-TDDFT simulations with HDFs. To this end, the key quantity to evaluate is the exact exchange component of the XC potential, which in the hybrid gauge is formally given by
\begin{align}
    \bar{V}^{\rm HF}_{{\rm x}, i j}(\boldsymbol{k}, t) & = \sum_{\boldsymbol{R}}  \,
    e^{i \boldsymbol{k} \cdot \boldsymbol{R}}
    \langle \bar{\phi}_{i {\bf 0}} \,|\, \tilde{V}^{\rm HF}_{{\rm x}}(\bfr,\bfrp, t)  \,|\, \bar{\phi}_{j \bfR} \rangle,
    \nonumber \\
    & = \sum_{\boldsymbol{R}}  \,
    e^{i \boldsymbol{k} \cdot \boldsymbol{R}}
    \langle \phi_{i {\bf 0}} \,|\, e^{i \boldsymbol{A}(t) \cdot (\bfr - \boldsymbol{\tau}_{i})}e^{-i \boldsymbol{A}(t) \cdot \bfr} V^{\rm HF}_{{\rm x}}(\bfr,\bfrp) e^{i \boldsymbol{A}(t) \cdot \bfrp} e^{-i \boldsymbol{A}(t) \cdot (\bfrp - \boldsymbol{\tau}_{j} - \bfR)} \,|\, \phi_{j \bfR} \rangle,
    \nonumber \\
    & = \sum_{\boldsymbol{R}}  \, 
    e^{i \boldsymbol{k} \cdot \boldsymbol{R}} e^{-i \boldsymbol{A}(t) \cdot \boldsymbol{\tau}_{i {\bf 0}, j \bfR}}
    \langle \phi_{i {\bf 0}} \,|\,  V^{\rm HF}_{{\rm x}}(\bfr,\bfrp)  \,|\, \phi_{j \bfR} \rangle,
    \label{eq:HFX_matrix_hybrid_gauge}
\end{align}
where 
\begin{align}
     V^{\rm HF}_{{\rm x}}(\bfr,\bfrp)&=K(\bfr-\bfrp)D(\bfr,\bfrp) \nonumber \\
     &= K(\bfr-\bfrp)\sum_{k,\bfR_1; l, \bfR_2}\phi_{k,\bfR_1}(\bfr)\bar{D}_{kl}(\bfR_2-\bfR_1)\phi_{l,\bfR_2}(\bfrp) \,
     \label{eq:HFX_potential}
\end{align}
with $K(\bfr-\bfrp)$ being the inter-electronic interaction kernel, and $D(\bfr,\bfrp)$ being the density matrix on real-space grids. Furthermore, 
  \begin{align}
      \bar{D}_{kl}(\bfR) 
      = \frac{1}{N_\bfk}\sum_{n,\bfk} e^{-i\bfk \cdot \bfR} f_{n\bfk}e^{i\boldsymbol{A}(t)\cdot (\boldsymbol{\tau}_k - \boldsymbol{\tau}_l - \bfR)}\bar{c}_{n\bfk}^k \bar{c}_{n\bfk}^{l\ast}  
     \label{eq:density_matrix}
  \end{align}
 is the density matrix within the NAO basis, where $k,l$ refers to the NAOs, $f_{n\boldsymbol{k}}$ is the occupation number. In hybrid-gauge calculations, the density matrix should be evaluated using Eq.~\ref{eq:density_matrix} since $\bar{c}_{n\bfk}^k$ are the KS eigenvectors that are directly accessible in such calculations.
 Here, the exact form of the interaction kernel $K(\bfr-\bfrp)$ depends
 on the actual HDFs: For full HFX, $K(\bfr-\bfrp)=v(|\bfr-\bfrp|)=1/|\bfr-\bfrp|$ is the bare Coulomb interaction, while for
 range-separated HDFs, it has a more flexible form, as will be detailed below.

 Putting Eqs.~\ref{eq:HFX_matrix_hybrid_gauge} and \ref{eq:HFX_potential} together, we obtain 
 \begin{align}
     \bar{V}^{\rm HF}_{{\rm x}, i j}(\boldsymbol{k}, t) 
    & = \sum_{\boldsymbol{R}}  \, 
    e^{i \boldsymbol{k} \cdot \boldsymbol{R}} e^{-i \boldsymbol{A}(t) \cdot \boldsymbol{\tau}_{i {\bf 0}, j \bfR}}
    \sum_{k,l,\bfR_1,\bfR_2} ( \phi_{i {\bf 0}} \phi_{k \bfR_1} \,|\, \phi_{l \bfR_2} \phi_{j \bfR} )
    D_{kl}(\bfR_2-\bfR_1),
    \label{eq:HFX_matrix_hybrid} \\
    & = \sum_{\boldsymbol{R}}  \, 
    e^{i \boldsymbol{k} \cdot \boldsymbol{R}} e^{-i \boldsymbol{A}(t) \cdot \boldsymbol{\tau}_{i {\bf 0}, j \bfR}}
    V_{{\rm x},ij}^{\rm HF}(\bfR)
    \label{eq:HFX_matrix}
 \end{align}
 where 
 \begin{equation}
    ( \phi_{i {\bf 0}} \phi_{k \bfR_1} \,|\, \phi_{l \bfR_2} \phi_{j \bfR} )
  =\iint d\bfr d\bfrp \phi_{i {\bf 0}}(\bfr) \phi_{k \bfR_1}(\bfr) K(\bfr-\bfrp) \phi_{l \bfR_2}(\bfrp) \phi_{j \bfR}(\bfrp)  
 \end{equation}
 are the two-electron interaction integrals.  Eq.~\ref{eq:HFX_matrix} indicates that the evaluation of
 the HFX matrix in the hybrid gauge can be done similarly as the case of usual HDF calculations, except that atomic-position-dependent
 phase factors need to be incorporated for each matrix element. This is in full consistency with Eq.~\ref{eq:Ham_hybrid_gauge}.

 Within the NAO basis framework, the HFX matrix can be efficiently calculated using the resolution of identity (RI)
 technique \cite{RI/Whitten/1973,RI/Dunlap/2010,RI/Vahtras/1993,RI/Ren/2012}, which  boils down to
 expanding the products of NAOs in terms of a set of auxiliary basis functions (ABFs),
 \begin{equation}
     \phi_{i {\bf 0}}(\bfr) \phi_{k \bfR_1}(\bfr)=\sum_{\mu,\bfR_2} C_{i{\bf 0},k \bfR_1}^{\mu \bfR_2} P_{\mu \bfR_2}(\bfr)
     \label{eq:RI_expan}
 \end{equation}
where $P_{\mu \bfR}(\bfr) = P_\mu(\bfr - \boldsymbol{\tau}_{\mu} -\bfR$) are atom-centered ABFs
and $C_{i{\bf 0},k \bfR_1}^{\mu \bfR_2}$ are the RI expansion coefficients. Using Eq.~\ref{eq:RI_expan},
the HFX matrix can be evaluated via
 \begin{align}
     V_{{\rm x},ij}^{\rm HF}(\bfR) & =\sum_{k,l,\bfR_1,\bfR_2}( \phi_{i {\bf 0}} \phi_{k \bfR_1} \,|\, \phi_{l \bfR_2} \phi_{j \bfR} )  D_{kl}(\bfR_2-\bfR_1) \nonumber  \\
     & = \sum_{k,l,\bfR_1,\bfR_2} \sum_{\mu,\nu, \bfR_3, \bfR_4} C_{i{\bf 0},k \bfR_1}^{\mu \bfR_3} 
     V_{\mu\bfR_3, \nu\bfR_4} C_{l\bfR_2,j\bfR}^{\mu \bfR_3} D_{kl}(\bfR_2-\bfR_1)
     \label{eq:V_HFX_RI}
 \end{align}
 where
 \begin{equation}
     V_{\mu\bfR_3, \nu\bfR_4}=\iint d\bfr d\bfrp P_{\mu\bfR_3}(\bfr) K(\bfr-\bfrp) P_{\nu\bfR_4}(\bfrp)
     \label{eq:Coulomb_matrix}
 \end{equation}
 is the so-called Coulomb matrix. Since $P_{\mu\bfR}(\bfr)$ are atom-centered functions, the elements of
 the Coulomb matrix in Eq.~\ref{eq:Coulomb_matrix} are two-center integrals, which can be efficiently computed in
 the reciprocal space,
 \begin{equation}
     V_{\mu\bfR_3, \nu\bfR_4}=\int d\bfq e^{-i\bfq \cdot (\bfR_4+{\boldsymbol \tau}_\nu-\bfR_3-{\boldsymbol \tau}_\mu)} \tilde{P}^\ast_{\mu}(\bfq) \tilde{K}(q) \tilde{P}_{\nu}(\bfq)
     \label{eq:Coulomb_matrix_kspace}
 \end{equation}
 where $\tilde{P}_{\mu}(\bfq)$ and $\tilde{K}(q)$ are the Fourier transforms of the ABF $P_{\mu}(\bfr)$ and 
 the interaction kernel $K(r)$, respectively. Note that the integration in Eq.~\ref{eq:Coulomb_matrix_kspace} goes over
 the entire reciprocal space instead of just the first BZ.

 In recent years, a localized RI (LRI) approach, where the ABFs are required to be located on the atoms where the two NAOs are centered \cite{RI/Ihrig/2015}, has been proposed. This approach has considerably boosted the efficiency of
 hybrid functional calculations \cite{RI/Levchenko/2015,ABACUS/Lin/2020,ABACUS/Lin/2021,ABACUS/Lin/2025,RI/Kokott/2024}, thanks to the resulting high sparsity of the RI expansion coefficients. This approach is also adopted in the present work.

\subsection{The RSH formalism and Coulomb singularity}
In the literature, many different flavors of HDFs have been developed, excelling in various territories of application. 
In the case of RSH functionals, the actual form of the HDFs depends crucially on the choice of the interaction kernel 
$K(\bfr - \bfrp)$. Specifically, the functional form can be conveniently specified via the Coulomb-attenuating method (CAM), where the bare
Coulomb operator $v(r)$ is range-partitioned using the complementary error function, $\erfc(x)$, as follows \cite{CAM/Yanai/2004}
\begin{equation}
    v(r)=\frac{1}{r}= v_1(r)+v_2(r)=\frac{\alpha+\beta \erfc(\mu r)}{r}
    +\frac{1-[\alpha+\beta \erfc({\mu r})]}{r},
    \label{eq:First:DE:cam}
\end{equation}
where the first term $v_1(r)=[\alpha+\beta \erfc(\mu r)]/r$ is used in the evaluation of the HFX contribution, and the second term $v_2(r)= 1/r - v_1(r)$ is incorporated into the evaluation of the KS semilocal exchange. 
In Eq.~\ref{eq:First:DE:cam}, $\alpha$ and $\alpha+\beta$ represent the fractions of the long- and short-range HFX components, respectively, and $\mu$ is the range-separation parameter. 
Based on this separation, the HFX contribution is evaluated via Eqs.~\ref{eq:V_HFX_RI} and \ref{eq:Coulomb_matrix} by setting $K(\bfr-\bfrp)=v_1(|\bfr-\bfrp|)$.
Then the entire exchange energy of the RSH functional can be expressed as
\begin{equation}
        E^\text{RSH}_\text{x} = \alpha E^\text{HF-LR}_\text{x}+(\alpha+\beta) E^\text{HF-SR}_\text{x}+(1-\alpha)E^\text{KS-LR}_\text{x}+[1-(\alpha+\beta)]E^\text{KS-SR}_\text{x}.
    \label{eq:First:DE:cam_func}
\end{equation}
 
 The decay behavior of the RSH exchange potential in real space is governed by the above-mentioned three parameters: $\alpha, \beta$, and  $\mu$. In the case of short-ranged corrected hybrid (SRCH) functionals---such as the HSE functional \cite{HSE/Heyd/2003,HSE/Heyd/2006} (with parameters $\alpha=0$, $\beta=0.25$, $\mu=0.106$ Bohr$^{-1}$)---the operator $v_1(r)$ decays rapidly, and the corresponding Fourier-transformed potential $\tilde{v_1}(q)$ remains finite in the long-wavelength limit ($\bfq\to0$). In contrast, for the bare Coulomb potential $v_1(r)=1/r$ used in conventional HF and GH formulations, $\tilde{v_1}(q)$ diverges as $1/q^2$ as $\bfq\to 0$ for 3-dimensional systems. Generally speaking, for any RSH functionals with $\alpha \ne 0$, 
 $\tilde{v_1}(q)$ will show diverging behavior as $\bfq$ approaches the $\Gamma$ point. This gives rise to 
 an integrable divergence at $q=0$ in the calculation of exact-exchange energy in $\bfk$ space. Neglecting the
 $\Gamma$ point in the BZ integration incurs an error of $O(N_\bfk^{1/3})$, making the $\bfk$-point convergence to 
adequate accuracy impractical. In the present work, we adopted an RI-based real-space formalism for the HFX calculations, as discussed in Sec.~\ref{sec:HFX_matrix}. The Coulomb singularity problem is manifested in the Coulomb matrix introduced in Eq.~\ref{eq:Coulomb_matrix}, which contains matrix elements that exhibit a slow $1/|\bfR_3 - \bfR_4|$ decaying behavior. This necessitates an exceedingly large BvK supercell for the lattice summation in Eq.~\ref{eq:V_HFX_RI} to converge, which is unfeasible in practical calculations. 

In the literature, two types of approaches have been developed to deal with the Coulomb singularity problem: The cut Coulomb operator method and the auxiliary function correction method.  
In Ref.~\citenum{Singularity/Spencer/2008}, Spencer and Alavi proposed a truncated Coulomb potential to replace the original bare Coulomb potential as an effective way to eliminate the Coulomb singularity. In real space, it is expressed as the product of $1/r$ and a step-like cutoff function
\begin{equation}
    v_\text{cut}(r;R_c)=
    \begin{cases}
        \frac{1}{r}, ~~~~ 0\leq r< R_c \\
        0, ~~~~ r\geq R_c
    \end{cases},
    \label{eq:TDDFT:Rc}
\end{equation}
which translates to the following truncated form of the Coulomb potential in reciprocal space 
\begin{equation}
    \begin{aligned}
        \tilde{v}_\text{cut}(q)&=\frac{2\sqrt{2}}{q^2}\biggl\{\alpha\largebra{1-\cos(qR_c)}                                                                             \\
     &\quad\quad\quad\quad - 0.5\beta\biggl[-2+2\cos(qR_c)\erfc(\mu R_c)                                   \\
     &\quad\quad\quad\quad +e^{-\frac{q^2}{4\mu^2}}\biggl(\erf\left(\frac{\ci q+2\mu^2R_c}{2\mu}\right) \\
     &\quad\quad\quad\quad -\ci\operatorname{erfi}\left(\frac{q}{2\mu}+\ci\mu R_c\right)\biggr)
    \biggr]
    \biggr\}\, ,
    \end{aligned}
    \label{eq:TDDFT:rsh_rcut_kernel}
\end{equation}
with $\operatorname{erfi}(x)=-\ci \operatorname{erf}(\ci x)$ being the imaginary error function. The cutoff radius $R_c$ can be estimated from the volume of the BvK supercell
\begin{equation}
    R_c=(\frac{3N_\bfk\Omega}{4\pi})^\frac{1}{3},
    \label{eq:TDDFT:sRc}
\end{equation}
where $N_\bfk$, as defined previously, is the number of $\bfk$-points in the first BZ, and $\Omega$ denotes the volume of the primitive cell. Further generalizations of the Spencer–Alavi scheme have subsequently been proposed in various electronic structure methods \cite{GW/Ren/2021, Singularity/Sundararaman/2013}. Within the RI framework, the real-space Coulomb matrix $V(\bfR)$ (defined in \refeq{eq:Coulomb_matrix}) can be constructed via Eq.~\ref{eq:Coulomb_matrix_kspace} by setting $\tilde{K}(q) =\tilde{v}_\text{cut}(q)$  (\refeq{eq:TDDFT:rsh_rcut_kernel}). The Coulomb matrix computed in this way decays to zero at the boundary of
the BvK supercell, enabling efficient evaluation of the real-space HFX matrix (\refeq{eq:V_HFX_RI}). 

An alternative, \bfk-space-based approach for treating the integrable singularity is to add and subtract an auxiliary function $F(\bfq)$\cite{Singularity/Gygi/1986,Singularity/Massidda/1993,Singularity/Carrier/2007,Singularity/Broqvist/2009}. Specifically, the HFX matrix in reciprocal space can be schematically expressed as
\begin{equation}
\begin{aligned}
    V^\text{HF}_{\rm x}(\bfk) &= \frac{\Omega}{(2\pi)^3} \int_{\text{BZ}} \dd\bfq\, M(\bfk - \bfq)\, \tilde{v}(\bfq) \\
                      &= \frac{\Omega}{(2\pi)^3} \int_{\text{BZ}} \dd \bfq \, \left[ M(\bfk - \bfq)\, \tilde{v}(\bfq) - M(\bfk)\, F(\bfq) \right] \\
                      &\quad + M(\bfk)\, \frac{\Omega}{(2\pi)^3} \int_{\text{BZ}} \dd^3q\, F(\bfq) \\
                      &\approx \frac{1}{N_\bfk} \sum_{\bfq \neq 0}^{\text{BZ}} M(\bfk - \bfq)\, \tilde{v}(\bfq) \\
                      &\quad + \frac{1}{N_\bfk} M(\bfk) 
                      \underbrace{ \left[ \frac{N_\bfk \Omega}{(2\pi)^3} \int_{\text{BZ}} \dd^3q\, F(\bfq) 
                      - \sum_{\bfq \neq 0}^{\text{BZ}} F(\bfq) \right] }_{\varsigma}.
\end{aligned}
\label{eq:HFX_aux_corr}
\end{equation}
Equation~\ref{eq:HFX_aux_corr} can be obtained by inserting Eq.~\ref{eq:Coulomb_matrix_kspace} into Eq.~\ref{eq:V_HFX_RI} and setting the interaction kernel to the bare Coulomb interaction $\tilde{K}(q)=\tilde{v}(q) = 4\pi/q^2$.
Here, $\varsigma$ is a correction term that accounts for the contribution from the integrable singularity of the Coulomb matrix $\tilde{v}(\bfq)$ at $\bfq = 0$. The quantity  $M(\bfk - \bfq)$ contains all the information aside from the Coulomb interaction, such as the RI expansion coefficients and the density matrix. One of the most widely used forms of the $F(\bfq)$ is suggested by Massidda \textit{et al} in Ref.~\citenum{Singularity/Massidda/1993}, 
in which
\begin{equation}
    F(\bfq)=\sum_\bfG\frac{e^{-\gamma|\bfq+\bfG|^2}}{|\bfq+\bfG|^2},
    \label{eq:tddft:massidda} 
\end{equation}
where $\bfG$ denotes the reciprocal lattice vectors. The parameter $\gamma$ is chosen such that the width of the Gaussian is comparable to the diameter of the BZ. To efficiently evaluate $V^{\rm HF}_{\rm x}(\bfq)$, the RI framework can be combined with the Ewald summation\cite{Singularity/Ewald/1921}, which decomposes the bare Coulomb interaction into a short-range part to be evaluated in real space and a long-range part to be evaluated in reciprocal space (see SI). 

\subsection{Evaluation of the current density with nonlocal HFX}
The time-dependent current density $\pmb{j}(t)$ in the length gauge can be calculated as \cite{TDDFT/Pemmaraju/2019,TDDFT/Zhao/2025}
\begin{equation}
    \pmb{j}(t)=-\frac{1}{2\Omega N_\bfk}\sum_{n\bfk}f_{n\bfk}\largebra{\bracket{\psi_{n\bfk}}{\hat{\bfp}}{\psi_{n\bfk}}+c.c},
    \label{eq:tddft:current}
\end{equation}
where $f_{n\bfk}$ is the occupation number and $\hat{\bfp}$ is the generalized momentum operator given by 
\begin{equation}
   \hat{\bfp} = \frac{1}{\ci}[\hat{\bfr},\hat{H}^\text{GKS}]. \label{eq:gp}
\end{equation}
The GKS Hamiltonian $\hat{H}^\text{GKS}$ contains nonlocal terms that do not commute with the position operator $\hat{\bfr}$, including the kinetic energy operator, the nonlocal part of the pseudopotential $\hat{V}_\text{NL}^\text{ps}$ and the nonlocal HFX.  While $[\hat{\bfr}, \hat{V}_\text{NL}^\text{ps}]$ can be directly computed via time-consuming grid-point integration, the corresponding term $[\hat{\bfr}, \hat{\bar{V}}_\text{NL}^\text{ps}]$ in the hybrid gauge, with respect to the NAOs $\phi_{i {\bf 0}}$ and $\phi_{j \bfR}$, can be expressed as
\begin{align}
    [\hat{\bfr}, \hat{\bar{V}}_\text{NL}^\text{ps}]_{ij}
    & = \sum_{\boldsymbol{R}}  \, 
    e^{i \boldsymbol{k} \cdot \boldsymbol{R}} e^{-i \boldsymbol{A}(t) \cdot \boldsymbol{\tau}_{i {\bf 0}, j \bfR}}
    \langle \phi_{i {\bf 0}} \,|\,  [\hat{\bfr}, \hat{V}_\text{NL}^\text{ps}]  \,|\, \phi_{j \bfR} \rangle.
    \label{eq:rV_hybrid_gauge}
\end{align}
Thus, the current density in the hybrid gauge without considering the Fock exchange contribution can be obtained from
 \begin{align}
     \pmb{j}(t)&=-\frac{1}{\Omega N_\bfk}\sum_{n,\bfk} f_{n\bfk}\Re \largebra{\sum_{ij,\boldsymbol{R}}e^{i\bfk \cdot \bfR}  e^{-i \boldsymbol{A}(t) \cdot \boldsymbol{\tau}_{i {\bf 0}, j \bfR}}\bar{c}_{n\bfk}^k \bar{c}_{n\bfk}^{l\ast}\langle \phi_{i {\bf 0}} \,|\,  -\ci\nabla+\ci[\hat{\bfr}, \hat{V}_\text{NL}^\text{ps}]  \,|\, \phi_{j \bfR} \rangle  
     \label{eq:j_hybrid_gauge}}
 \end{align}
To include the HFX contribution in the hybrid gauge, one needs to evaluate $[\hat{\bfr}, \hat{V}_{\rm x}^\text{HF}]_{ij}$, which is formally given by
\begin{equation}
    \langle \phi_{i {\bf 0}} \,|\,  [\hat{\bfr}, \hat{V}_{\rm x}^\text{HF}]  \,|\, \phi_{j \bfR} \rangle = \frac{1}{N_\bfk}\sum_{\bfR_1\bfR_2}D(\bfR_2-\bfR_1)[ ( \phi_{i {\bf 0}} \phi_{k \bfR_1} \,|\bfr|\, \phi_{l \bfR_2} \phi_{j \bfR} ) - (  \phi_{l \bfR_2} \phi_{j \bfR}\,|\bfr|\,  \phi_{i {\bf 0}} \phi_{k \bfR_1} )], \label{eq:rV_HFX}
\end{equation}
where
\begin{equation}
    ( \phi_{i {\bf 0}} \phi_{k \bfR_1} \,|\bfr|\, \phi_{l \bfR_2} \phi_{j \bfR} ) =  \iint d\bfr d\bfrp \phi_{i {\bf 0}}(\bfr) \phi_{k \bfR_1}(\bfr) \bfr K(\bfr-\bfrp) \phi_{l \bfR_2}(\bfrp) \phi_{j \bfR}(\bfrp).
\end{equation}
However, evaluating \refeq{eq:rV_HFX} is challenging in periodic systems due to its lack of translational invariance. To address this, we insert the identity operator $I = \sum_{ij}\ket{\phi_i}S_{ij}^{-1}\bra{\phi_j}$ into \refeq{eq:gp}, where $S_{ij}=\bracketw{\phi_i}{\phi_j}$ denotes the overlap matrix of the non-orthogonal basis. 
As a consequence, the generalized momentum matrix can be calculated within the AO basis \cite{PYATB/Jin/2021,PYATB/Jin/2023} as
\begin{equation}
    \begin{aligned}
        p_{ij,a}(\bfk)&=\partial_a H^\text{GKS}_{ij}(\bfk)+\ci\largebra{\sum_{kl}H^\text{GKS}_{ik}(\bfk)S^{-1}_{kl}(\bfk)A^R_{lj,a}(\bfk)-A^R_{ik,a}(\bfk)S^{-1}_{kl}(\bfk)H^\text{GKS}_{lj}(\bfk)} \\
        &\quad-\sum_{kl}H^\text{GKS}_{ik}(\bfk)S^{-1}_{kl}(\bfk)\partial_a S_{lj}(\bfk), 
    \end{aligned}
    \label{eq:gm}
\end{equation}
where
\begin{equation}
    \begin{aligned}
        H^\text{GKS}_{ij}(\bfk)&=\sum_{\bfR}e^{\ci\bfk\cdot\bfR}\bracket{\phi_{i0}}{\hat{H}}{\phi_{j\bfR}},\\
        S_{ij}(\bfk)&=\sum_{\bfR}e^{\ci\bfk\cdot\bfR}\bracketw{\phi_{i 0}}{\phi_{j \bfR}},\\
        A^R_{ij,a}(\bfk)&=\sum_{\bfR}e^{\ci\bfk\cdot\bfR}\bracket{\phi_{i 0}}{\bfr_a}{\phi_{j \bfR}},\\
        \partial_a H_{ij}(\bfk) &= \ci\sum_{\bfR}R_ae^{\ci\bfk\cdot\bfR}\bracket{\phi_{i 0}}{\hat{H}}{\phi_{j \bfR}},\\
        \partial_a S_{i
        j}(\bfk) &= \ci\sum_{\bfR}R_ae^{\ci\bfk\cdot\bfR}\bracketw{\phi_{i 0}}{\phi_{j \bfR}}, \label{eq:gm_detail}
    \end{aligned}
\end{equation}
with $a=x,y,z$. By substituting $\psi_{n\bfk}(\bfr,t)=\sum_{i}c_{ni\bfk}(t)\phi_{i\bfk}(\bfr)$ into \refeq{eq:tddft:current} and utilizing the relation $p_{ij,a}=\bra{\phi_{i\bfk}(\bfr)}\hat{\bfp}_{a}\ket{\phi_{j\bfk}(\bfr)}$, the time-dependent current density can be obtained as
\begin{equation}
    j_a(t) = \Re \largebra{\sum_{\bfk}\sum_{ij} D_{ij}(\bfk, t) \, p_{ji,a}(\bfk, t) }, \label{eq:J}
\end{equation}
with the single-particle density matrix $D_{ij}(\bfk,t)=\sum_n f_{nk}c_{n\bfk}^i(t)c^{j\ast}_{n\bfk}(t)$. It is worthwhile to note that the computational scheme presented in \refeq{eq:gm}-(\ref{eq:J}) is applicable to arbitrary Hamiltonian. This derivation is equally valid in the velocity gauge. In the case of the hybrid gauge, $\hat{\bar{H}}$ and $\hat{\bar{S}}$ are needed in \refeq{eq:gm_detail}, and the corresponding current density can be obtained by substituting $D$ with $\bar{D}$. Unlike the approach in Ref.~\citenum{TDDFT/Pemmaraju/2018}, which computes $[\hat{\bfr}, \hat{\tilde{V}}_{nl}]$ separately using time-consuming grid integrals, this method relies solely on matrix multiplication, leading to improved efficiency. The optical response function is obtained from the time-dependent current,
\begin{equation}
    \sigma_{ab}(\omega)=\frac{J_a(\omega)}{E_b(\omega)}=\frac{\int_0^T \dd t e^{\ci\omega t }J_a(t)}{\int_0^T \dd t e^{\ci\omega t }E_b(t)}\, ,
\end{equation}
and the dielectric function is then given by
\begin{equation}
    \varepsilon(\omega) = \varepsilon_0 + \ci\frac{\sigma(\omega)}{\omega},
\end{equation}
where the imaginary part, $\Im\varepsilon(\omega)$, characterizes the optical absorption.

\section{Result}
We first compare the Spencer–Alavi cut Coulomb operator and auxiliary function correction methods for treating the Coulomb singularity across different hybrid functionals in bulk systems such as Si, AlP, and NaCl. 
Subsequently, we apply RT-TDDFT-RSH using different gauges to simulate exciton dynamics in periodic systems. The RT-TDDFT-RSH framework discussed above is implemented in the Atomic-orbital Based Ab-initio Computation at USTC (ABACUS) package \cite{ABACUS/Chen/2010,ABACUS/Li/2016,ABACUS/Ji/2022}. All calculations employed second-generation numerical atomic orbital (NAO) basis sets, specifically the DPSI sets \cite{ABACUS/Lin/2021/dpsi}, in conjunction with SG15 \cite{Pseudopotential/schlipf/2015} optimized norm-conserving Vanderbilt-type pseudopotentials \cite{Pseudopotential/vanderbilt/1990}.

\subsection{Assessing Coulomb singularity schemes}
\begin{figure}[!htbp]
	\includegraphics[width=0.8\textwidth,trim=0cm 0cm 0cm 0cm]{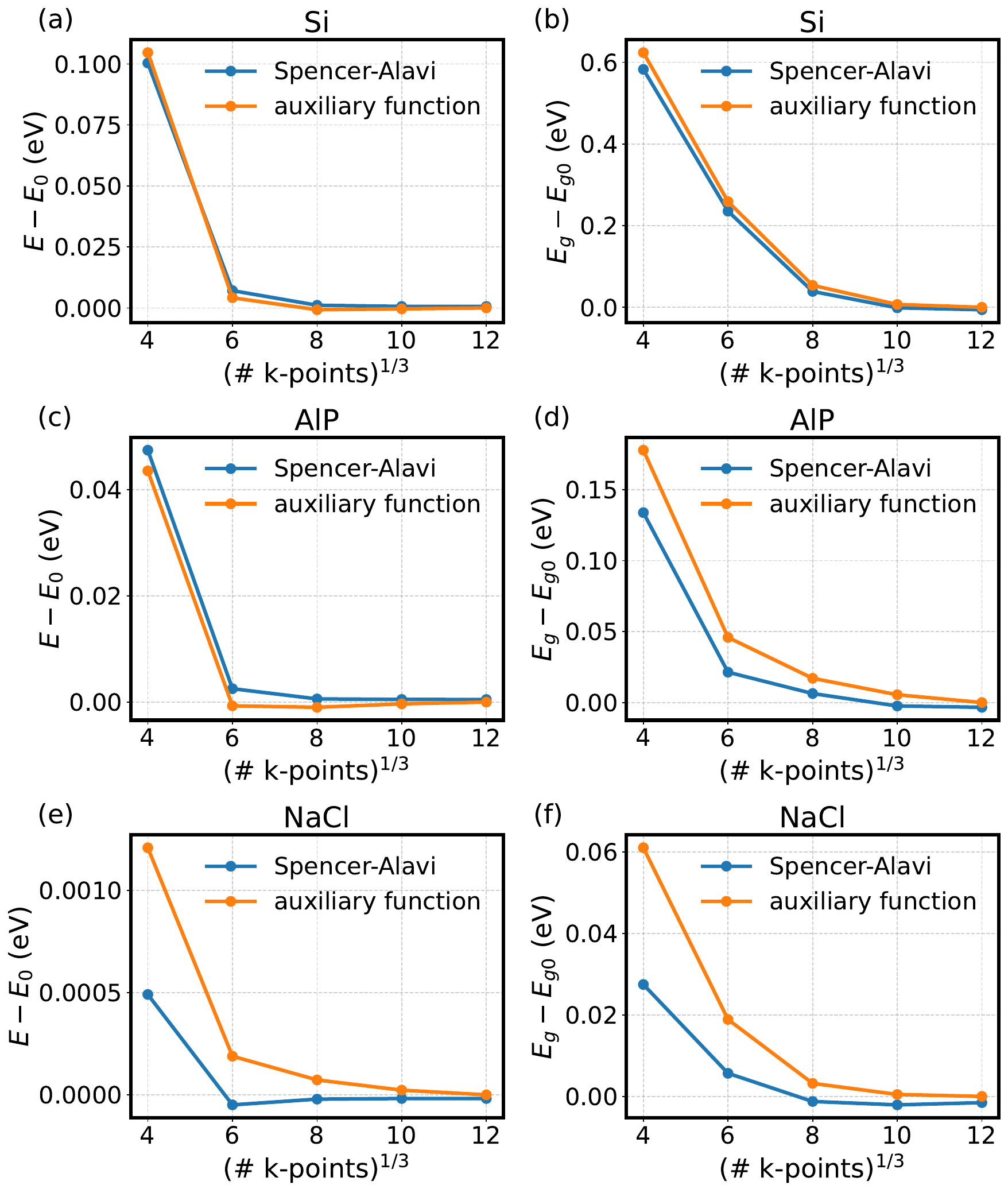}
	\centering
    \caption{HF calculations using the Spencer-Alavi and the auxiliary function correction methods are performed to obtain the energy (on the left) and band gap (on the right) variations with respect to the $\bfk$-points for Si, AlP, and NaCl. The results are referenced to those obtained with the auxiliary function correction method, using $\bfk$-points set to $12 \times 12 \times 12$ ($E_0$/$E_{g0}$). 
	\label{fig:tddft:hf_com}}
\end{figure}

In \reffig{fig:tddft:hf_com}, the variations in energy and band gap for Si, AlP, and NaCl with respect to the number of $\bfk$-points in the BZ sampling are shown, as obtained from HF calculations using the Spencer-Alavi and the auxiliary function correction methods.  As the number of $\bfk$-points increases, the energies and band gaps of the three semiconductors/insulators gradually converge to the same values. The $\bfk$-point convergence behaviors of the Spencer-Alavi and auxiliary function correction methods are fairly similar, with only minor differences. 

\begin{figure}[!htbp]
	\includegraphics[width=0.8\textwidth,trim=0cm 0cm 0cm 0cm]{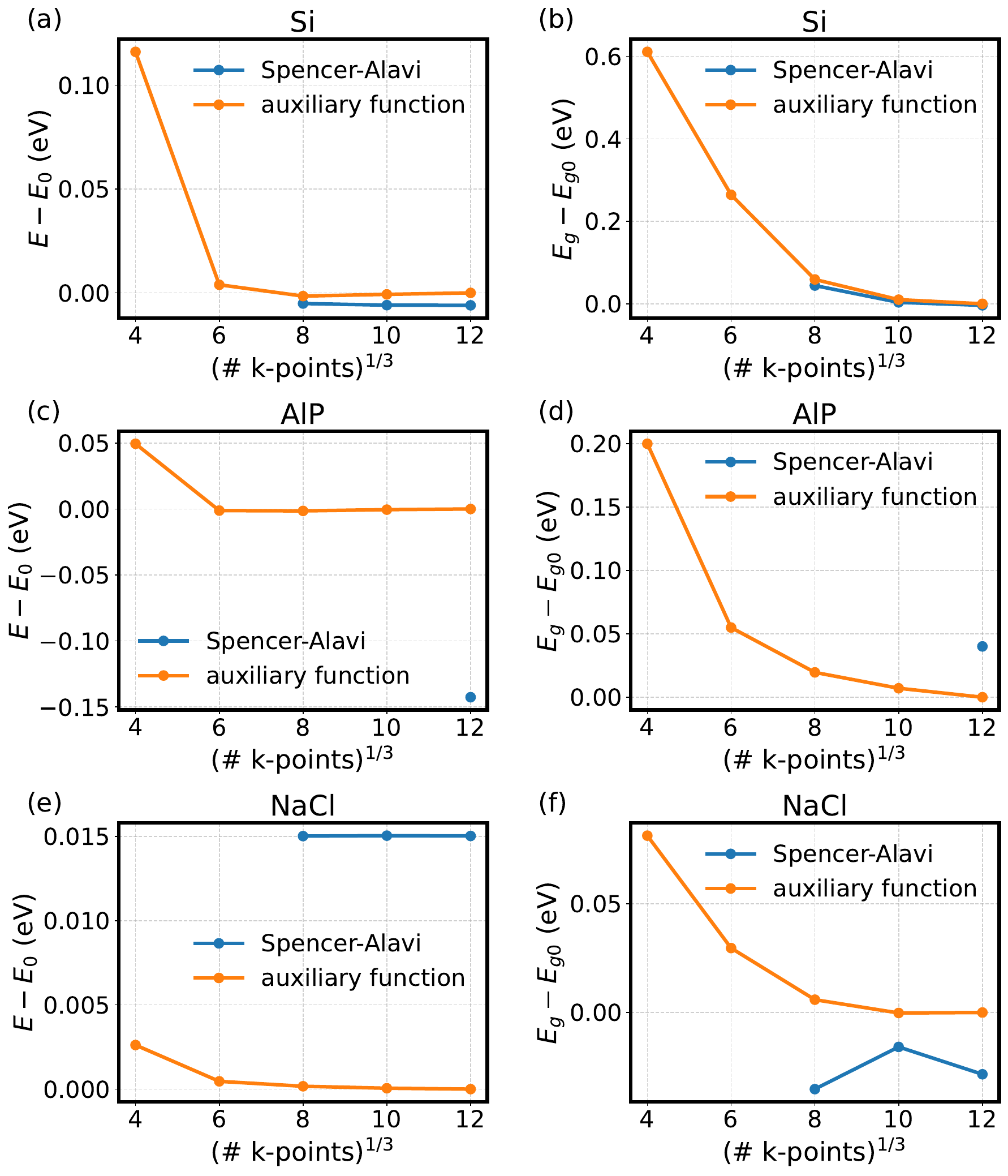}
	\centering
    \caption{LC-PBE calculations using the Spencer-Alavi and the auxiliary function correction methods are performed to obtain the energy (on the left) and band gap (on the right) variations with respect to the $\bfk$-points for Si, AlP, and NaCl. The results are referenced to those obtained with the auxiliary function correction method, using $\bfk$-points set to $12 \times 12 \times 12$ ($E_0$/$g_0$). When the number of $\bfk$-points is insufficient, the Spencer-Alavi method calculations do not converge, and therefore no data points are shown in the figure. 
	\label{fig:tddft:lc_pbe_com}}
\end{figure}

However, for the long-range corrected PBE (LC-PBE) functional \cite{LCPBE/Tawada/2004}, with mixing parameters $\alpha = 1.0$ and $\beta = -1.0$, and a screening parameter $\mu = 0.33$ Bohr$^{-1}$, the Spencer-Alavi method provides stable energy and band gap predictions only when the number of $\bfk$-points exceeds a certain limit, i.e., when the cutoff radius (as defined in \refeq{eq:TDDFT:Rc}) is adequately large such that the truncated Coulomb potential can still effectively capture the long-range behavior of the bare Coulomb potential. In contrast, the auxiliary function correction method ensures consistent convergence of both energies and band gaps as the number of $\bfk$-points increases (see \reffig{fig:tddft:lc_pbe_com}).

\begin{figure}[!htbp]
	\includegraphics[width=0.6\textwidth,trim=0cm 0cm 0cm 0cm]{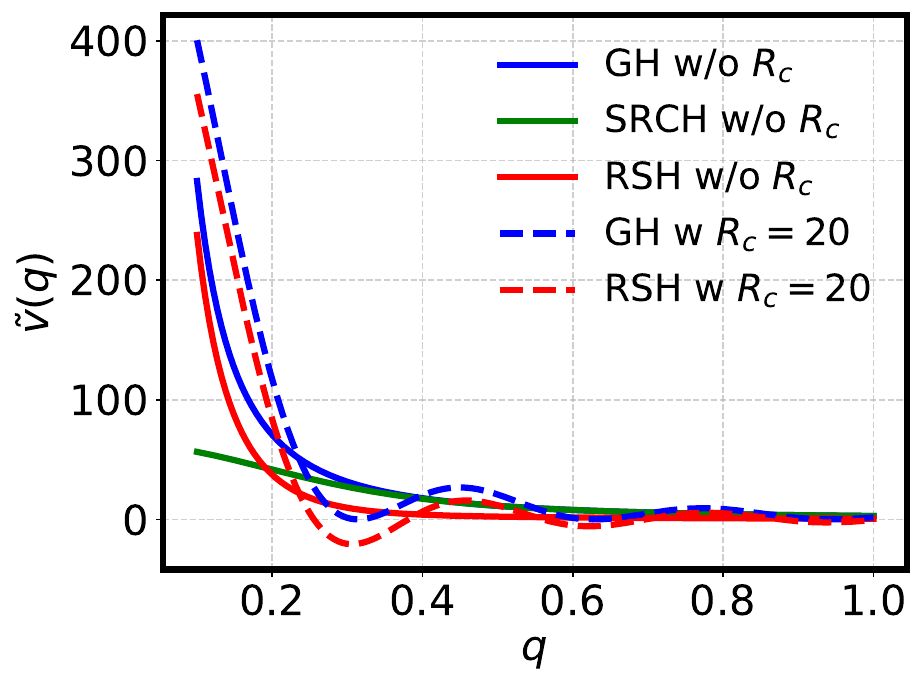}
	\centering
    \caption{The variation of the radial functions of different Coulomb potentials with respect to $q$ is shown, where GH refers to the bare Coulomb potential ($\alpha = 1.0$, $\beta = 0.0$), SRCH to the short-range Coulomb potential ($\mu = 0.106$ Bohr$^{-1}$, $\alpha = 0.0$, $\beta = 1.0$), and RSH to the range-separated Coulomb potential ($\mu = 0.106$ Bohr$^{-1}$, $\alpha = 1.0$, $\beta = -0.8$), corresponding to the blue, green, and red solid lines, respectively. The blue and red dashed lines represent the truncated bare Coulomb potential and range-separated Coulomb potential at $R_c = 20$ Bohr, respectively. 
    \label{fig:tddft:kernel}}
\end{figure}

To gain insights into what is really going on, we compare different forms of the interaction kernel $v_1(r)$ (Eq.~\ref{eq:First:DE:cam}) used in the HFX calculations in reciprocal space. In \reffig{fig:tddft:kernel}, we plot the Fourier-transformed $\tilde{v}_1(q)$ of five types of
interaction kernels, including  the short-ranged potential ($v^\text{SRCH}(r)=\frac{\text{erfc}(\mu r)}{r}$), the bare potential ($v^\text{GH}(r)=\frac{1}{r}$),  the range-separated 
potential ($v^\text{RSH}(r)=\frac{\alpha+\beta\text{erfc}(\mu r)}{r}$), as well as the truncated forms of the latter two. 
The issue with the truncated form of the range-separated Coulomb potential for describing long-range inter-electronic interactions becomes evident in reciprocal space. As can be seen in \reffig{fig:tddft:kernel}, compared to SRCH, both GH and the long-range dominant RSH decay more slowly, as expected. When a real-space truncation radius is introduced, the truncated form of GH exhibits oscillatory decaying behavior but remains positive. In contrast, the truncated form of RSH oscillates around zero, leading to a non-positive-definite Coulomb matrix. This causes numerical problems and can destroy the stability of the calculations.

\subsection{RT-TDDFT-RSH simulations of exciton dynamics}
According to \refeq{eq:First:DE:cam}, the RSH functionals require additional parameters to describe the exchange interactions, including $\alpha$, $\beta$ and $\mu$. Choosing proper values for these parameters has long been a challenge in hybrid functional theory research. The long-range mixing parameter $\alpha$ is typically chosen to accurately describe the asymptotic behavior of the Coulomb interaction in a dielectric medium, which decays as $\frac{1}{\epsilon_\infty r}$, where $\epsilon_\infty$ represents the macroscopic dielectric constant\cite{CAM/Alkauskas/2011}. This value can be obtained from experiments or first-principles calculations \cite{CAM/Chen/2018,CAM/Yang/2023,TDDFT/Refaely/2015,TDDFT/Pemmaraju/2019,CAM/Yang/2023}. For three-dimensional materials, the short-range contributions controlled by $\alpha + \beta$ are generally set between 0.20 and 0.25. In contrast, for two-dimensional materials, whose long-wavelength dielectric constant tends to unity \cite{GW/Falco/2013},  setting $\epsilon_\infty$ to 1 is effective, which implies $\alpha=1$ and $\beta=0$, meaning that only 100\% long-range HFX is considered. Finally, $\mu$ can be estimated based on experimental or $GW$ band gaps. The parameter settings for the benchmark studies in the present work are provided in \reftab{tab:tddft:settings}. 


\begin{table}[ht]
    \centering
    \caption{The mixing coefficients ($\alpha$ and $\beta$) and the screening parameter ($\mu$, in units of bohr$^{-1}$) used in the absorption simulations for three functionals \cite{TDDFT/Pemmaraju/2019}: SRCH: short-range corrected hybrid functional, LRCH: long-range corrected hybrid functional, GH: global hybrid functional. \label{tab:tddft:settings}}
    \begin{tabular}{c|c|c}
         \toprule
             & Si  & 2D h-BN \\
        \midrule
        SRCH & $\alpha=0.0$& \\
         & $\beta=0.25$ & --\\
         & $\mu=0.11$& \\
        LRCH & $\alpha=0.0833$  & $\alpha=1$\\
         & $\beta=0.1167$ & $\beta=-1$\\
         & $\mu=0.11$&$\mu=0.126$ \\
        GH &  & $\alpha=0.5$\\
        & --  & $\beta=0.0$\\
        &   & $\mu=0.0$\\
         \bottomrule
    \end{tabular}
\end{table}

\begin{figure}[!htbp]
	\includegraphics[width=\textwidth,trim=0cm 0cm 0cm 0cm]{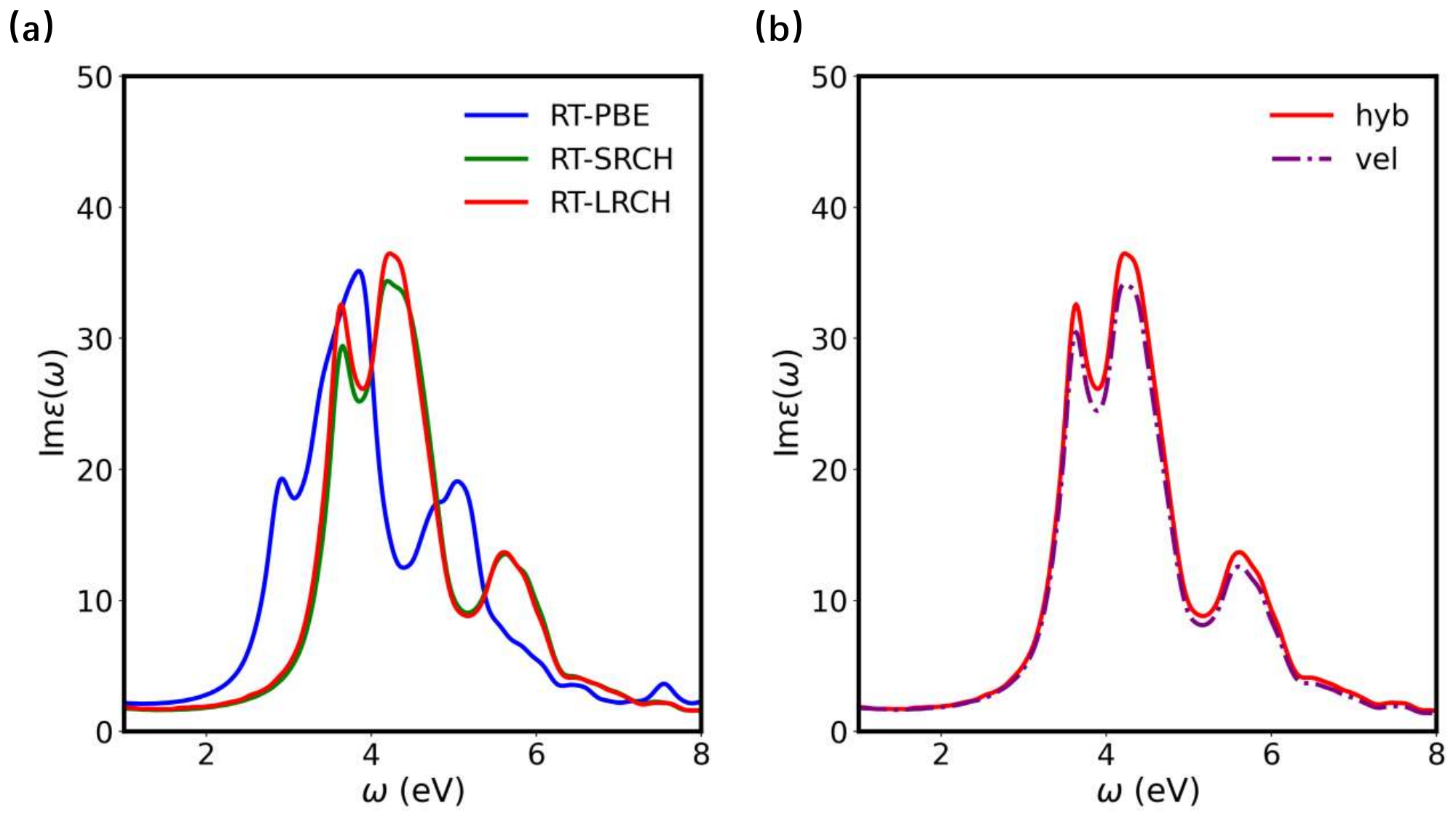}
	\centering
    \caption{(a) Absorption spectra ($\Im\varepsilon(\omega)$) of Si obtained from RT-TDDFT calculations using different functionals (PBE, SRCH, LRCH), with simulations performed using the hybrid gauge and the external electric field applied along the \(z\)-direction.  (b) A comparison of the Si absorption spectra obtained from RT-LRCH simulations using the hybrid (hyb) and velocity (vel) gauges.
	\label{fig:tddft:abs}}
\end{figure}

In \reffig{fig:tddft:abs}(a), we compare the absorption spectra of the three-dimensional semiconductor Si obtained from RT-TDDFT simulations using the hybrid gauge with the PBE, short-range corrected hybrid (SRCH), and long-range corrected hybrid (LRCH) functionals. The SRCH functional employed in the present work, with the chosen parameters listed in \reftab{tab:tddft:settings}, corresponds to the renowned HSE functional \cite{HSE/Heyd/2003}. 
The simulations employed a double-$\zeta$ basis set plus polarization (DZP) functions, specifically [2$s$2$p$1$d$] for Si, and a $ 10 \times 10 \times 10 $ $\bfk$-point grid.  A 0.052 $V$/\AA~ delta-function electric field pulse was applied at $t = 0.0048$ fs to induce a time-dependent current density, from which the imaginary part of the dielectric function ($\Im\varepsilon(\omega)$) was computed to characterize the system’s linear optical response. A time step of 0.0024 fs was used throughout the simulations, with a total propagation time of 24 fs.

RT-PBE yields absorption spectra with peak positions at lower energies, whereas both RT-SRCH and RT-LRCH induce a blue shift. Both RT-SRCH and RT-LRCH can capture the excitonic absorption peak near 3.5 eV, while RT-LRCH produces a more pronounced absorption feature. The differences between the hybrid and velocity gauges in RT-LRCH simulations are further illustrated in \ref{fig:tddft:abs}(b). Because the velocity gauge does not account for position-dependent phase information, the resulting absorption peaks are significantly weaker than those obtained with the hybrid gauge. A detailed comparison of results obtained with the velocity, hybrid, and length gauges for finite systems at the PBE level has been presented in \refcite{TDDFT/Zhao/2025}. We extend this comparison to the level of hybrid functionals for C$_2$H$_4$ (see Figure. S1 and S2). The current densities obtained with the length and hybrid gauges are nearly identical, whereas that obtained with the velocity gauge exhibits deviations. 

Compared to its three-dimensional counterpart, 2D h-BN exhibits a weaker screening effect, resulting in a stronger exciton binding energy \cite{BSE/Thygesen/2017}. In the RT-TDDFT simulation of monolayer h-BN, we constructed a vacuum layer with a thickness of 30 \AA~ along the $z$-direction and used a $24 \times 24 \times 1$ \bfk-point grid. For both B and N elements, the DZP basis settings ([2$s$2$p$1$d$]) were employed. The external electric field was applied as a delta-function pulse with an intensity of 0.052 V/Å, oriented along the in-plane polarization direction of h-BN. The total simulation time was 24 fs for RT-PBE, whereas for RT-LRCH and RT-GH a longer duration of 60 fs was employed in order to obtain well-resolved absorption peaks.

\begin{figure}[!htbp]
	\includegraphics[width=\textwidth,trim=0cm 0cm 0cm 0cm]{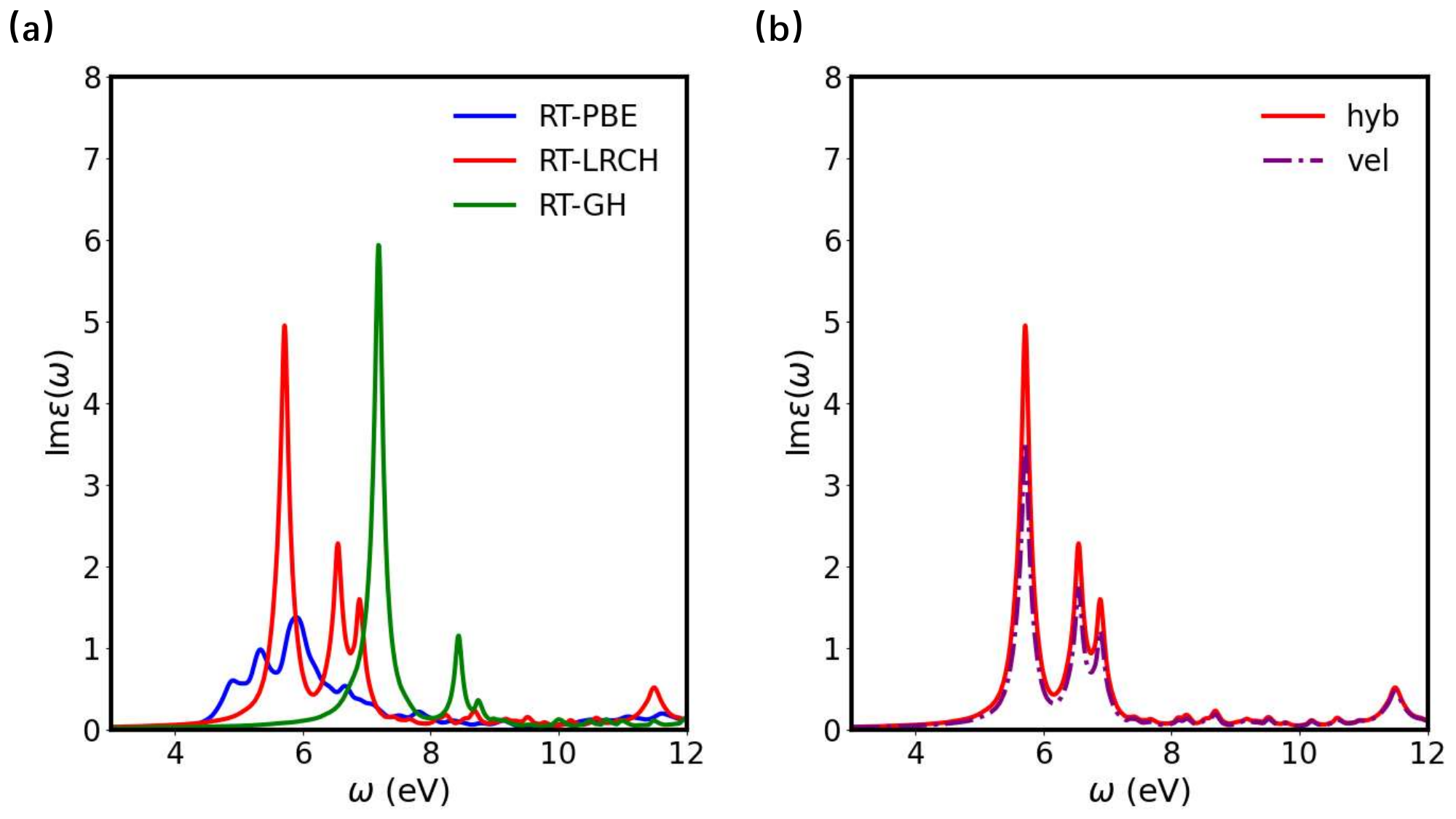}
	\centering
    \caption{The current density $j(t)$ and the imaginary part of the dielectric function $\Im\varepsilon(\omega)$ for h-BN calculated using RT-TDDFT with different functionals (PBE, LRCH, GH), are shown.
	\label{fig:tddft:hbn-abs}}
\end{figure}

As shown in \reffig{fig:tddft:hbn-abs}(a), compared to the RT-PBE results, the RT-LRCH absorption spectrum shows a sharper excitonic absorption peak around 5.8 eV and captures a smaller absorption peak near 7.0 eV. Compared to RI-LRCH, the RT-GH results exhibit a further blue shift due to the overestimated band gap. Additionally, we compare the absorption spectra of h-BN obtained from RT-LRCH simulations under the hybrid and velocity gauges. As shown in \reffig{fig:tddft:hbn-abs}(b), the hybrid gauge provides a more accurate description of the excitonic effects, while the velocity gauge significantly underestimates the excitonic absorption peak due to its inherent  errors.

\begin{figure}[!htbp]
	\includegraphics[width=1.0\textwidth,trim=0cm 0cm 0cm 0cm]{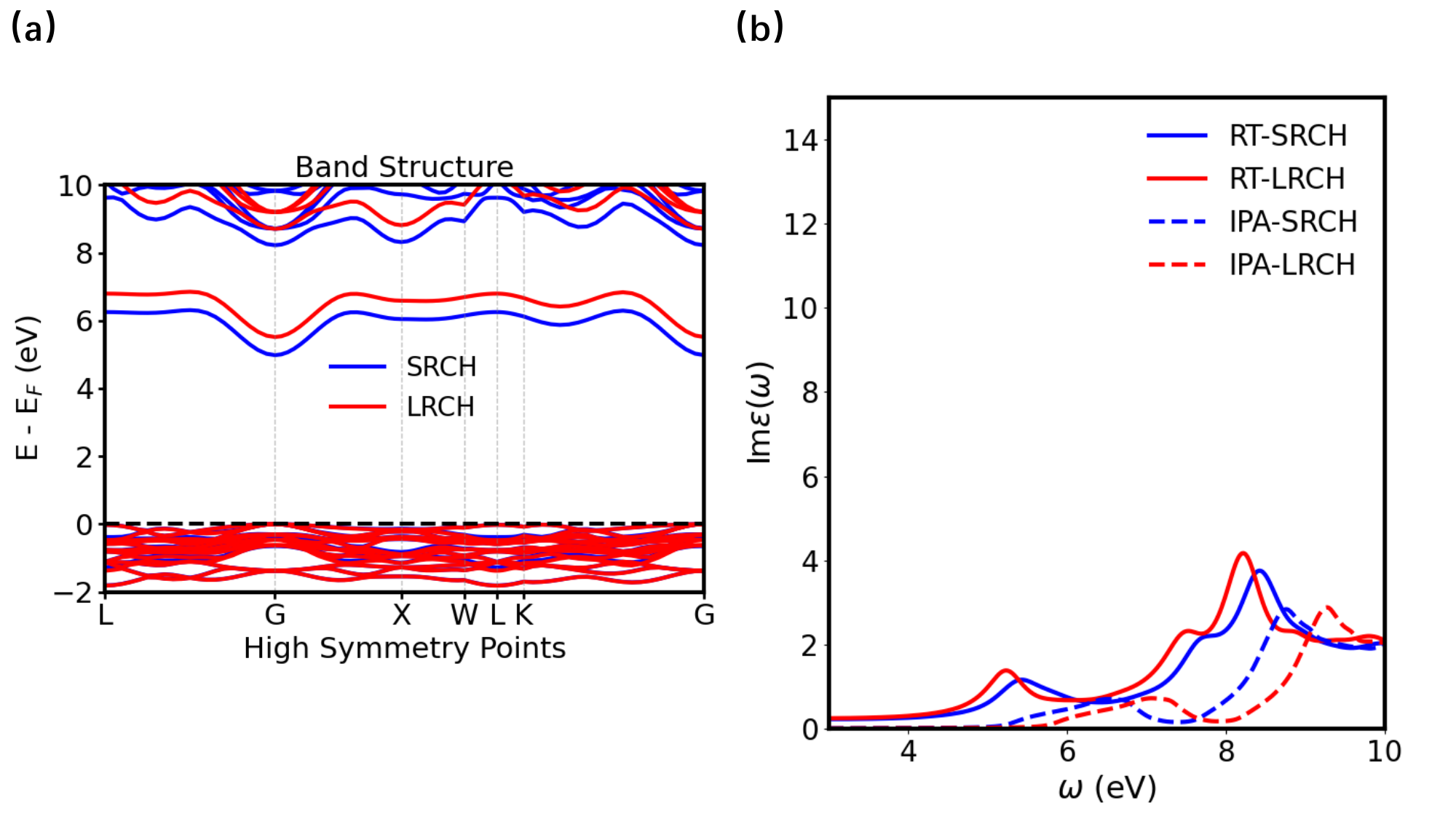}
	\centering
    \caption{(a) Band structure of Cs$_2$NaInCl$_6$ obtained from static calculations using the SRCH and LRCH methods. (b) Absorption spectra of Cs$_2$NaInCl$_6$ simulated via the RT-SRCH and RT-LRCH methods (solid lines), compared with the IPA-SRCH and IPA-LRCH results (dashed lines). 
	\label{fig:tddft:hdp}}
\end{figure}

The double perovskite Cs$_2$NaInCl$_6$ exhibits a large fundamental bandgap, which significantly deviates from the experimental optical gap due to the prominent exciton effects\cite{HDP/Biega/2023,HDP/Ji/2024,HDP/Linn/2024}. In this study, we compare the differences between RT-SRCH and RT-LRCH in predicting the excited-state properties of Cs$_2$NaInCl$_6$. For SRCH, the parameters $\alpha=0.0$, $\beta=0.4$ and $\mu=0.11$ bohr$^{-1}$ are used, which have been validated to yield a more reasonable ground-state structure for double perovskites\cite{HDP/Ji/2024}. While for LRCH, the long-range contribution is inferred to be 36\% ($\alpha=0.36$) from the dielectric constant (2.78) obtained from RPA\cite{HDP/Biega/2023}, with $\beta=-0.16$, and $\mu=0.28$ bohr$^{-1}$ set to match the $G_0W_0$@PBE bandgap (5.52 eV). The ground-state band structures obtained from SRCH and LRCH static calculations with DZP basis sets, namely, Cs with [4s2p1d], In with [2s2p2d], Cl with [2s2p1d], Na with [4s2p1d], are shown in \reffig{fig:tddft:hdp}(a). The bandgap calculated with SRCH is 4.98 eV, slightly smaller than the 5.52 eV obtained with LRCH. The dispersion is similar for both methods, leading to only a small difference in the effective masses.

In \reffig{fig:tddft:hdp}(b), we compare the optical absorption spectra of Cs$_2$NaInCl$_6$ simulated using the RT-LRCH and RT-SRCH methods. With the inclusion of long-range corrections, the absorption peak around $\sim 5.4$ eV in the RT-LRCH spectrum becomes more pronounced compared to the RT-SRCH results. However, unlike other double perovskite materials, such as Bi-based double perovskites, the overall excitonic absorption remains relatively weak. This is due to the fact that band-edge transitions in In-based perovskites are dipole-forbidden \cite{HDP/Biega/2023,HDP/Luo/2018}. The band gap inferred from the absorption spectrum based on the independent-particle approximation is consistent with the fundamental band gap obtained from the corresponding band structure in \reffig{fig:tddft:hdp}(a). Compared to the IPA results, the RT-LRCH absorption spectrum exhibits a more pronounced redshift, while the RT-SRCH redshift is relatively smaller. This suggests that RT-LRCH provides a more accurate description of excitonic effects, whereas RT-SRCH tends to underestimate the exciton binding energy.

\section{Summary}
This study implements the Spencer-Alavi and auxiliary function correction methods for RSH functional calculations based on NAO basis sets. Benchmarking on three-dimensional materials reveals that, compared to the commonly used Spencer-Alavi truncated potential, the auxiliary function corrected Ewald summation ensures computational completeness by separating the calculations in real space and reciprocal space. Specifically, when the RSH functional involves significant long-range exchange interactions, the Spencer-Alavi method may lead to non-positive-definite Coulomb matrices, whereas the the auxiliary function corrected Ewald summation effectively mitigates this issue.


Integrating the RSH functional with RT-TDDFT enables a more precise description of excitonic effects in materials. However, in the velocity-gauge RT-TDDFT-RSH simulations with NAOs, the absence of position-dependent phase information can lead to an underestimation of excitonic absorption peaks. To overcome this limitation, this study expands the time-dependent Bloch wavefunctions using a hybrid-gauge basis that incorporates phase information. By carefully optimizing the mixing and screening parameters of the RSH functional, we significantly enhance the accuracy of RT-TDDFT in capturing excitonic effects. Additionally, since directly computing the contribution of nonlocal HFX to the current density is highly challenging, we employ a completeness relation approach to enable efficient and precise current density calculations. Through excited-state simulations of Si, monolayer h-BN, and Cs$_2$NaInCl$_6$, we demonstrate that our work establishes a more accurate RT-TDDFT-RSH implementation that better predicts excitonic effects in periodic systems.



\section*{Acknowledgments}
We acknowledge the funding support from the National Natural Science Foundation of China (Grants Nos.
12134012, 12374067, and 12188101) and the Strategic Priority Research Program of the Chinese Academy of Sciences 
(Grant No. XDB0500201). This work was also funded by the National Key Research and Development Program of China (Grant Nos. 2022YFA1403800 and 2023YFA1507004) and the robotic AI-Scientist platform of the Chinese Academy of Sciences.

\begin{suppinfo}
The following file is available free of charge.
\begin{itemize}
  \item supporting\_info.pdf:
  The file contains the following items: the Ewald summation for evaluating Coulomb matrix and RT-TDDFT-RSH simulation for C$_2$H$_4$.
  \end{itemize}
\end{suppinfo}

\bibliography{comm}

\end{document}


\renewcommand{\thepage}{S\arabic{page}}
\renewcommand{\thefigure}{S\arabic{figure}}
\renewcommand{\thetable}{S\arabic{table}}
\renewcommand{\theequation}{S\arabic{equation}}

\section{The Ewald summation for evaluating Coulomb matrix}
The Ewald summation scheme \cite{Singularity/Ewald/1921}, commonly employed for evaluating the Hartree energy, decomposes the bare Coulomb interaction into two parts: a short-range component treated in real space and a long-range component handled in reciprocal space. An analogous strategy can be adopted for computing HFX integrals in reciprocal space. Specifically, the discrete Fourier transform of the Coulomb matrix $V_{\mu\bfR_3,\nu\bfR_4}$ (Eq.~20 in the main text) is defined as
\begin{equation}
    \tilde{V}_{\mu\nu}(\bfq)=\sum_{\bfR_3,\bfR_4} V_{\mu\bfR_3,\nu\bfR_4}e^{\ci\bfq\cdot(\bfR_4-\bfR_3)}.
    \label{eq:tddft:Vq_c}
\end{equation}
Based on Eq.~\ref{eq:tddft:Vq_c} and Eq.~20, the summation over reciprocal lattice vectors $\bfG$ provides another defination for computing $\tilde{V}_{A\alpha,B\beta}(\bfq)$
\begin{equation}
        \tilde{V}_{\mu\nu}(\bfq) = \sum_{\bfG} \tilde{P}^\ast_{\mu}(\bfk) \tilde{K}(k) \tilde{P}_{\nu}(\bfk)e^{-\ci\bfk\cdot({\boldsymbol \tau}_\nu-{\boldsymbol \tau}_\mu)}\Bigg|_{\bfk=\bfq-\bfG}.
    \label{eq:tddft:Vq_d}
\end{equation}
When the multipole moments $p$ of $P$ are zero, $V_\bfR$ vanishes for $R$ beyond the sum of the effective radii of the charge distributions. Consequently, $\tilde{V}(\bfq)$ can be directly computed using \refeq{eq:tddft:Vq_c}. However, in practical calculations, since the NAO basis functions for ABFs possess finite multipole moments, the Ewald method is required to precisely compute the short-range and long-range Coulomb interactions between atom-centered ABFs in real (\refeq{eq:tddft:Vq_c}) and reciprocal space (\refeq{eq:tddft:Vq_d}), respectively.

By employing a set of Gaussians $\{f_{lm}(\bfr)\}$, where $l$ and $m$ denote the azimuthal and magnetic quantum numbers, respectively, and normalized them to unit multipole moments, ABFs can be transformed into multipole-free functions $P^{'}_\mu=P_\mu-p_\mu f_{l_\mu m_\mu}$ and $P^{'}_\nu=P_\nu-p_\nu f_{l_\nu m_\nu}$. With this transformation, the interaction $(P_{A\alpha}|P_{B\beta})$ can be expressed as
\begin{equation}
    (P_{\mu}|P_{\nu})=p_\mu p_\nu (f_{l_\mu m_\mu}|f_{l_\nu m_\nu})+(P_\mu|P^{'}_\nu)+(P^{'}_{\mu}|P_{\nu})-(P^{'}_{\mu}|P^{'}_{\nu}).
\end{equation}
Here, the last three terms can be directly computed using \refeq{eq:tddft:Vq_c}, while the first term, which represents the Gaussian interaction, is described by \refeq{eq:tddft:Vq_d}.

The Gaussian basis function $f_{lm}(\bfr)$ is defined in real and reciprocal space as
\begin{equation}
    f_{lm}(\bfr)=N_lr^le^{-\gamma r^2/2}Y_l^m(\hat{r}), \quad
    \tilde{f}_{lm}(\bfk)=\ci^{-l}\tilde{N}_lk^le^{-k^2/2\gamma}\tilde{Y}_l^m(\hat{k}).
\end{equation}
Applying the normalization coefficients
\begin{equation}
    N_l=\frac{\gamma^{l+3/2}}{\sqrt{\pi/2}(2l-1)!!},\quad
    \tilde{N}_l=\frac{N_l}{\gamma^{l+3/2}}=\frac{1}{\sqrt{\pi/2}(2l-1)!!},
\end{equation}
the unit multipole moment is obtained as
\begin{equation}
    \begin{aligned}
        p_f&=\frac{1}{2l+1}\int^{\infty}_0\dd r r^{2+l}f_l(r)\\
            &=\sqrt{\pi/2}(2l-1)!!\cdot\lim_{k\to0}\frac{\tilde{f}_l(k)}{k^l} =1,
    \end{aligned}
\end{equation}
where $f_l(r)$ and $\tilde{f}_l(k)$ denote the radial components of $f_{lm}(\bfr)$ and $\tilde{f}_{lm}(\bfk)$, respectively. The analytical expression for  $(f_{l_am_a}|f_{l_bm_b})$ is given by
\begin{equation}
    \begin{aligned}
        (f_{l_am_a}|f_{l_bm_b})=\sum_{LM}A_{LM}\sum_{\bfG}\largebra{k^{l_a+l_b-L}G_{LM}(\bfk)\tilde{v}(\bfk)e^{-\ci\bfk\cdot\tau}}\Bigg|_{\bfk=\bfq-\bfG},
    \end{aligned}
    \label{eq:tddft:Vq_f}
\end{equation}
where 
\begin{equation}
        A_{LM}=\ci^{(l_a-l_b)}\tilde{N}_{l_a}^*\tilde{N}_{l_b}C^{LM}_{l_am_al_bm_b},\quad
        G_{LM}(\bfk)=k^Le^{-\frac{k^2}{\gamma}}\tilde{Y}_{L}^{M}(\hat{k}),
\end{equation}
and the relation $\tilde{Y}_l^m(\hat{k})\tilde{Y}_{l'}^{m'}(\hat{k})=\sum_{LM}C^{LM}_{lm,l'm'}\tilde{Y}_{L}^{M}(\hat{k})$ is used. 

\section{RT-TDDFT-RSH simulations for C$_2$H$_4$}

\begin{figure}[!htbp]
	\includegraphics[width=1.0\textwidth,trim=3cm 0cm 3cm 0cm]{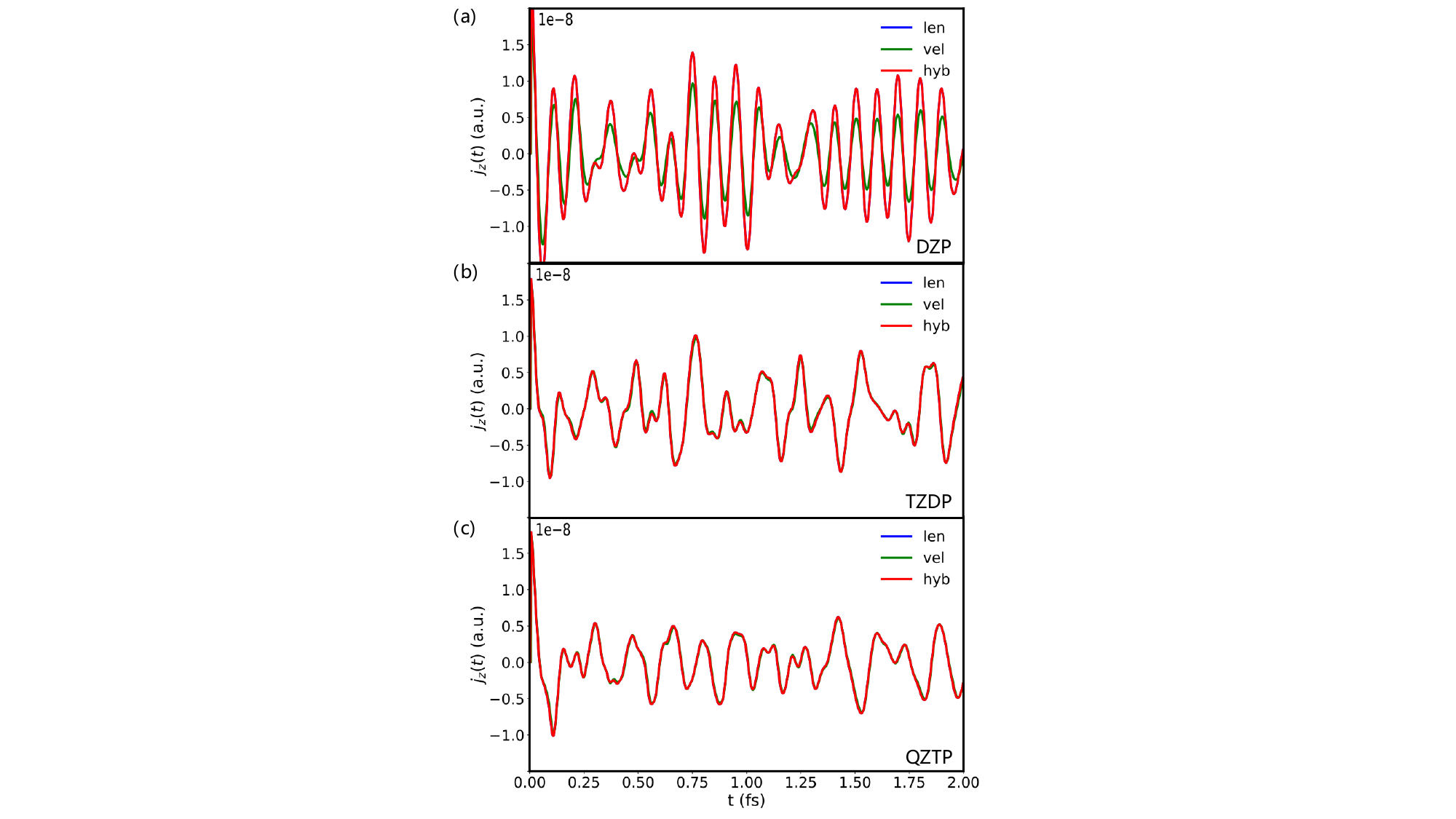}
	\centering
    \caption{RT-TDDFT simulations on the C$_2$H$_4$ molecule using the LC-PBE functional under the length, velocity, and hybrid gauges, illustrating current density based on (a) DZP, (b) TZDP and (c) QZTP basis sets.
	\label{fig:tddft:c2h4-current}}
\end{figure}

We performed RT-TDDFT simulations on the C$_2$H$_4$ molecule using the LC-PBE functional, comparing current densities and absorption spectrum obtained under the length, velocity and hybrid gauges and different basis sets, including DZP, TZDP and QZTP. A 0.052 $V$/\AA~ delta-function electric field pulse was applied at $t = 0.0048$ fs to induce a time-dependent current density, from which the imaginary part of the dielectric function ($\Im\varepsilon(\omega)$) is computed to characterize the system’s linear optical response. A time step of 0.0024 fs was used throughout the simulation. 

\begin{figure}[!htbp]
	\includegraphics[width=1.0\textwidth,trim=3cm 0cm 3cm 0cm]{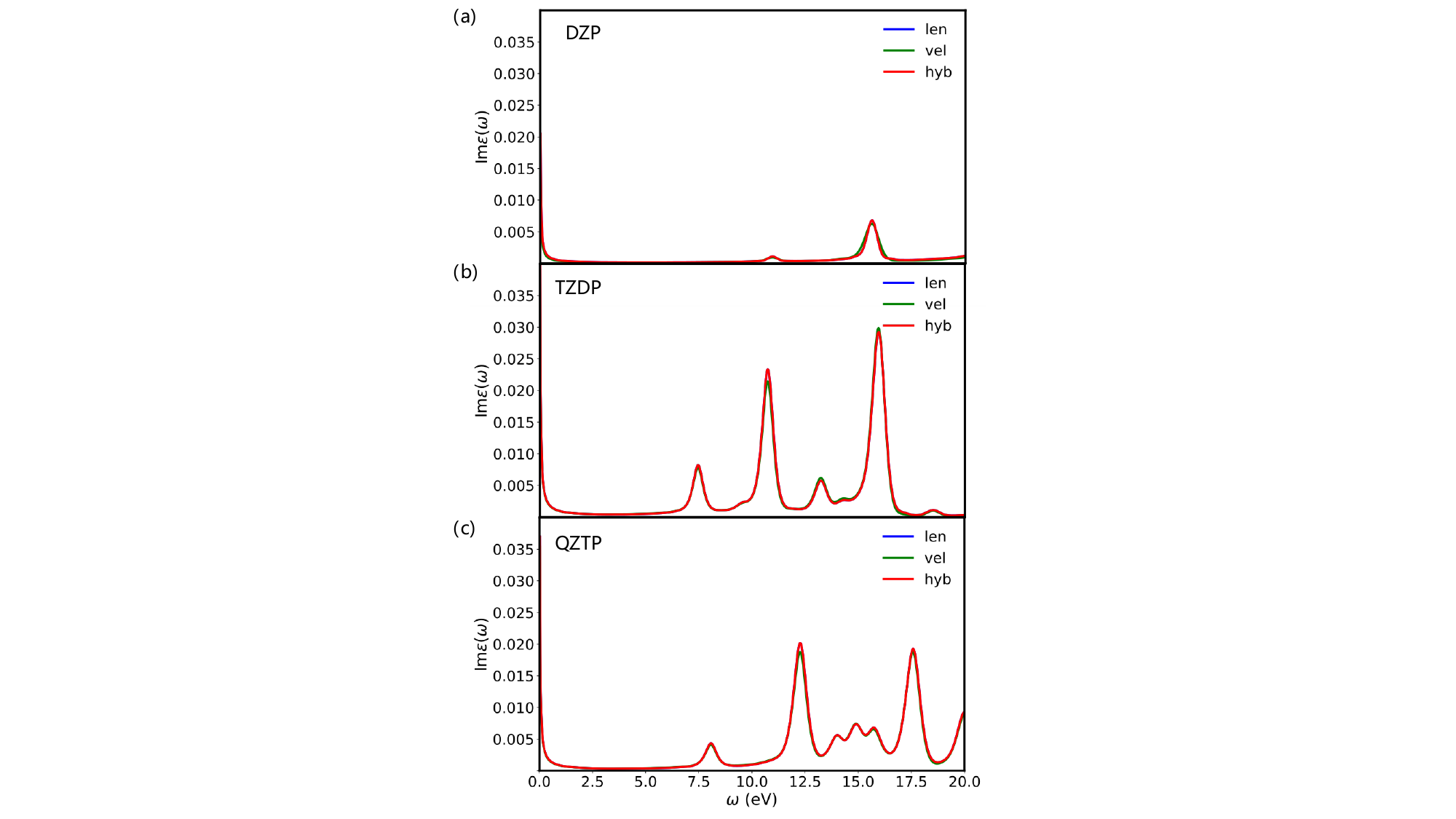}
	\centering
    \caption{RT-TDDFT simulations on the C$_2$H$_4$ molecule using the LC-PBE functional under the length, velocity, and hybrid gauges, illustrating absorption spectrum based on (a) DZP, (b) TZDP and (c) QZTP basis sets.
	\label{fig:tddft:c2h4-imag}}
\end{figure}

As shown in \reffig{fig:tddft:c2h4-current} and \ref{fig:tddft:c2h4-imag}, the current densities obtained with the length and hybrid gauges are nearly identical, whereas the velocity gauge exhibits deviations, leading to discrepancies in the absorption spectrum. At the hybrid functional level, this finding aligns with previous RT-TDDFT studies based on LDA/GGA functionals, reaffirming that the velocity gauge introduces unphysical artifacts in the absorption spectrum and the deviations decrease with increasing basis set size.\cite{TDDFT/Zhao/2025}. 

\bibliography{comm}

%% file: newcommands.tex
\newcommand{\ci}{\mathrm{i}}

\newcommand{\bfr}{ {\bf r}} 
\newcommand{\bfrp}{ {\bf r'}} 
\newcommand{\bfR}{ {\bf R}} 
\newcommand{\bfq}{ {\bf q}} 
\newcommand{\bfp}{ {\bf p}} 
\newcommand{\bfk}{ {\bf k}} 
\newcommand{\bfG}{ {\bf G}} 

\newcommand{\bracketw}[2]{\ensuremath{\langle #1 | #2  \rangle}}
\newcommand{\bracket}[3]{\ensuremath{\langle #1 | #2 | #3 \rangle}}
\newcommand{\ket}[1]{\ensuremath{| #1 \rangle}}

\newcommand{\dd}[1]{\text{d}#1}
\newcommand{\refeq}[1]{Eq.~\ref{#1}} 
\newcommand{\refcite}[1]{Ref.~\citenum{#1} }
\newcommand{\reffig}[1]{Figure.~\ref{#1} } 
\newcommand{\reftab}[1]{Table.~\ref{#1} }
\newcommand{\largebra}[1]{\left[#1\right]}

\renewcommand{\Re}{\operatorname{Re}}
\renewcommand{\Im}{\operatorname{Im}}
\def\bra#1{\mathinner{\langle{#1}|}}
\def\ket#1{\mathinner{|{#1}\rangle}}

\newcommand{\RNum}[1]{\uppercase\expandafter{\romannumeral #1\relax}}

\DeclareMathOperator\erf{erf}
\DeclareMathOperator\erfc{erfc}